\documentclass{emulateapj}

\shorttitle{Cosmic Optical Background}
\shortauthors{Matsuoka et al.}

\begin{document}


\title{Cosmic Optical Background: the View from Pioneer 10/11}


\author{Y. Matsuoka\altaffilmark{1}, N. Ienaka\altaffilmark{2}, K. Kawara\altaffilmark{2}, and S. Oyabu\altaffilmark{1}}


\altaffiltext{1}{Graduate School of Science, Nagoya University, Furo-cho, Chikusa-ku, Nagoya 464-8602, Japan; 
  matsuoka@a.phys.nagoya-u.ac.jp}
\altaffiltext{2}{Institute of Astronomy, The University of Tokyo, Osawa 2-21-1, Mitaka, Tokyo 181-0015, Japan}


\begin{abstract}
We present the new constraints on the cosmic optical background (COB) obtained from an analysis of the {\it Pioneer 10/11} 
Imaging Photopolarimeter (IPP) data.
After careful examination of data quality, the usable measurements free from the zodiacal light are integrated 
into sky maps at the blue ($\sim$0.44 $\mu$m) and red ($\sim$0.64 $\mu$m) bands.
Accurate starlight subtraction is achieved by referring to all-sky star catalogs and a Galactic stellar
population synthesis model down to 32.0 mag.
We find that the residual light is separated into two components: one component shows a clear correlation
with thermal 100 $\mu$m brightness, while another betrays a constant level in the lowest 100 $\mu$m brightness region.
Presence of the second component is significant after all the uncertainties and possible residual light in the Galaxy
are taken into account, thus it most likely has the extragalactic origin (i.e., the COB).
The derived COB brightness is (1.8 $\pm$ 0.9) $\times$ $10^{-9}$ and (1.2 $\pm$ 0.9) $\times$ $10^{-9}$ erg s$^{-1}$ cm$^{-2}$ 
sr$^{-1}$ \AA$^{-1}$ at the blue and red band, respectively, or 7.9 $\pm$ 4.0 and 7.7 $\pm$ 5.8 nW m$^{-2}$ sr$^{-1}$.
Based on a comparison with the integrated brightness of galaxies, we conclude that the bulk of the COB is comprised of 
normal galaxies which have already been resolved by the current deepest observations.
There seems to be little room for contributions of other populations including "first stars" at these wavelengths.
On the other hand, the first component of the IPP residual light represents the diffuse Galactic light (DGL)---scattered 
starlight by the interstellar dust.
We derive the mean DGL-to-100 $\mu$m brightness ratios of $2.1 \times 10^{-3}$ and $4.6 \times 10^{-3}$ at the two bands, 
which are roughly consistent with the previous observations toward denser dust regions.
Extended red emission in the diffuse interstellar medium is also confirmed.
\end{abstract}





\section{Introduction \label{sec:intro}}

The cosmic optical background (COB) is the optical component of the extragalactic background light,
which is the integrated radiation from all light sources outside the Galaxy.
Dominant contribution to the COB comes from stellar nucleosynthesis in galaxies at redshifts
$z < 10$, while other mechanisms such as mass accretion to super massive black holes in active 
galactic nuclei (AGNs), gravitational collapse of stars, and particle decay can contribute to the
COB brightness.
As a fossil record of light production activity in the Universe, the COB conveys information on the
cosmic star formation history including birth and death of Population III stars.
Comparison of the measured COB brightness and the integrated brightness of resolved galaxies has 
fundamental importance, since it provides a direct clue as to the fraction of celestial visible
light whose origin is known to us.

However, robust detection of the COB has long been hampered by the extremely bright foreground emissions.
While expected brightness of the COB is around 1 bgu $\equiv$ $1 \times 10^{-9}$ erg s$^{-1}$ cm$^{-2}$ 
sr$^{-1}$ \AA$^{-1}$, the terrestrial airglow and the zodiacal light (ZL) are a few orders of magnitude
brighter than this level at optical wavelengths \citep{leinert98}.
Besides these brightest two components, the diffuse Galactic light (DGL), which refers to scattered
starlight by the interstellar dust, is another but much fainter component of the diffuse light of the night sky.
\citet{dube77,dube79} attempt a COB measurement by subtracting all the foreground components
from the night-sky brightness observed at Kitt Peak National Observatory.
The terrestrial airglow is assumed to follow a function of the zenith angle with temporal variation, while
the ZL brightness is estimated from equivalent width of an imprinted solar Fraunhofer line on the ZL spectrum, 
making use of the fact that the solar spectrum is well reproduced in the ZL due to weak wavelength dependence 
in scattering strength of the interplanetary dust.
\citet{mattila76} devised the method to utilize Galactic dark clouds as shielding walls of the COB.
By comparing the brightness measured at two lines of sight, one toward a dark cloud and one toward
a nearby "transparent" sky, the background light coming from behind the dark cloud could be extracted.
However, these and other similar attempts were not very successful due to inadequate knowledge about the 
foreground radiations, especially spatial and temporal variations of their brightness.

One of the fundamental solutions to the problem of foreground subtraction is to carry out measurements at outside 
the relevant emission regions.
\citet{toller83} utilizes the measurements of the night-sky brightness by a photopolarimeter on board 
the {\it Pioneer 10} spacecraft.
He analyzes the data obtained at the heliocentric distances larger than 3.3 AU, where the ZL is found 
to be negligible \citep{hanner74}.
Unfortunately, though, Galactic stars cannot be resolved due to poor spatial resolution of the instrument, 
which dominate the measured sky brightness and prevent robust detection of the COB.

The latest attempt is made by \citet{bernstein02a,bernstein02b,bernstein02c}, who carried out a coordinated observation 
program using the {\it Hubble Space Telescope} ({\it HST}) and a ground-based telescope at Las Campanas Observatory.
The terrestrial airglow is avoided by using space observations, while the ZL brightness is estimated by a similar method 
to \citet{dube77,dube79}, i.e., from measured equivalent width of a solar Fraunhofer line reproduced in the ZL.
The DGL brightness is estimated using a simple scattering model and thermal emission brightness of the interstellar dust.
However, a number of problems in their analysis are raised by \citet{mattila03}, e.g., inappropriate assumptions in deriving 
the ZL and DGL brightness.
A series of corrections to the original papers are presented in \citet{bernstein05} and \citet{bernstein07},
and the derived COB brightness is now accompanied by significant errors;
the latest values are (6 $\pm$ 4), (10 $\pm$ 5), and (7 $\pm$ 4) bgu at the 0.30, 0.55, and 0.80 $\mu$m band, respectively.

Here we present the new constraints on the COB brightness obtained from an original analysis of the night-sky data taken by 
the Imaging Photopolarimeters (IPPs) on board the {\it Pioneer 10} and {\it 11} spacecrafts.
The present work gives a significant improvement of the previous analysis by \citet{toller83} by taking advantage of 
the additional higher-quality ({\it Pioneer 11}) data, accurate all-sky star catalogs, and a realistic DGL model combined 
with the far-infrared (IR) space observations.
This paper is organized as follows.
In Section 2 we describe the observations and reductions of the data used in this work, including data assessment 
and map creation processes.
Subtraction of the Galactic light, namely, the starlight and the DGL, are presented in Section 3.
Then we show the results in Section 4, followed by discussion on the derived correlation between the DGL and the far-IR 
emission and on the resolved fractions of the COB in Section 5.
Finally a summary appears in Section 6.
Hereafter the {\it Pioneer 10} and {\it 11} spacecrafts are sometimes abbreviated as "P10" and "P11".
All the magnitudes are given in the Vega-based system.
%We also use the abbreviated energy unit 'bgu', where 1 bgu = $10^{-9}$  erg s$^{-1}$ cm$^{-2}$ \AA$^{-1}$ sr$^{-1}$.

%We use the $S_{10}(V)$ units, the equivalent number of $V$ = 10 mag stars of solar color per square degree, 
%for expressing the brightness measured by the Pioneer IPPs.
%It can be converted to the bgu units approximately by 
%\begin{eqnarray*}
%  S_{10}(V) = 1.18 \times 10^{-9}\ {\rm erg} {\rm s}^{-1} {\rm cm}^{-2} {\rm \AA}^{-1} {\rm sr}^{-1}
%\end{eqnarray*} 
%for the IPP blue band and by
%\begin{eqnarray*}
%  S_{10}(V) = 1.07 \times 10^{-9}\ {\rm erg} {\rm s}^{-1} {\rm cm}^{-2} {\rm \AA}^{-1} {\rm sr}^{-1}
%\end{eqnarray*} 
%for the IPP red band \citep{weinberg81}.

\section{Observations and Reductions \label{sec:data}}

%% In a manner similar to \objectname authors can provide links to dataset
%% hosted at participating data centers via the \dataset{} command.  The
%% second curly bracket argument is printed in the text while the first
%% parentheses argument serves as the valid data set identifier.  Large
%% lists of data set are best provided in a table (see Table 3 for an example).
%% Valid data set identifiers should be obtained from the data center that
%% is currently hosting the data.
%%
%% Note that AASTeX interprets everything between the curly braces in the 
%% macro as regular text, so any special characters, e.g. "#" or "_," must be 
%% preceded by a backslash. Otherwise, you will get a LaTeX error when you 
%% compile your manuscript.  Special characters do not 
%% need to be escaped in the optional, square-bracket argument.

Here we describe the observations and reductions used in this work.
While a similar analysis is carried out by \citet{gordon98} for the investigation of the Galactic diffuse emission,
we clarify the two major differences between their and the present analysis as follows.
(1) Stellar contribution is removed from the {\it each} instrument field of view (FOV) before creating the final, high-resolution sky maps 
in this work, rather than first creating star-contaminated maps and then evaluating the stellar contribution in them.
Considering the large instrument FOV (low angular resolution), the former method is more suitable than the latter to derive the spatial 
distribution of the diffuse emission as we will see below.
(2) The data quality is evaluated at a few stages in the middle of the data processing in this work, which allows us the 
accurate understanding of the uncertainties in the final results.

\subsection{IPP Observations \label{sec:IPPobs}}

The {\it Pioneer 10} and its sister ship {\it Pioneer 11} are NASA's space missions, which were launched on 1972 March 2 
and 1973 April 5, respectively.
They are the first spacecrafts to travel through the Asteroid belt and explore the outer solar system.
One of the scientific instruments on board the spacecrafts is the IPP, which was designed to make
two-color (blue and red) maps of the brightness and polarization over the sky during cruise phase of the missions \citep{pellicori73}.
The IPP data taken at the heliocentric distances $R$ = 1 -- 5 AU were used to study distribution of the interplanetary dust 
responsible for the ZL and diffuse light components outside the ZL clouds \citep[e.g.,][]{hanner74, weinberg74, toller87}.
Below we give a description of the IPP measurements and the primary data processing.
For more details, see, e.g., \citet{weinberg74} and \citet{gordon98}.

The IPP is equipped with a 2.5 cm Maksutov-type telescope, together with a Wollaston prism, multilayer filters, and two 
dual-channel Bendix channeltrons \citep{pellicori73}.
It measures simultaneously two orthogonal polarization components of brightness in two (blue at the wavelengths 0.39 -- 0.50 $\mu$m 
and red at 0.60 -- 0.72 $\mu$m; "$B_{\rm IPP}$" and "$R_{\rm IPP}$" hereafter) broad-band filters.
Effective wavelengths of the two bands are 0.44 $\mu$m and 0.64 $\mu$m \citep{gordon98}.
The two polarization components are defined with "S" vector parallel to the instrument rotation axis and "P" vector
which is perpendicular to it.
Thus one IPP exposure generates four surface brightness measurements, $S^{\rm S}$ (S-vector component) and $S^{\rm P}$ 
(P-vector component) at each of the $B_{\rm IPP}$ and $R_{\rm IPP}$ band.
The four measurements are carried out by the different photocathodes placed at the end of the different optical paths.
% The IPP can be operated in three different modes.
The data used in the present work was obtained with the "ZL mode", with instantaneous FOV of 
2$^{\circ}$.29 $\times$ 2$^{\circ}$.29.
Sky brightness is integrated for 1/64 (one "sector") of a 12.5 s spacecraft spin period, resulting in effective 
FOV of 2$^{\circ}$.29 $\times$ (2$^{\circ}$.29 $+$ 5$^{\circ}$.625 sin$L$) where $L$ is the angle between the instrument 
pointing and the spacecraft spin axis (called "look angle").
With a change of look angle, the effective FOV ranges from 7.5 deg$^2$ ($L$ = 170$^{\circ}$) to 18 deg$^2$ ($L$ = 90$^{\circ}$).
Since the single-pixel detector is used in each of the four channels, the FOV of the detector size also gives the angular resolution
of the instrument.
A data cycle of the IPP measurements consists of 10 rolls; eight for sky measurements, one for photometric calibration 
using a self-luminous phosphor source, and one for monitoring offset and dark-current levels.
Sky maps are created by changing look angle $L$ and repeating the data cycles.

Primary processing of the obtained data has been carried out by the IPP team.
Initially absolute photometric calibration is achieved using the pre-flight measurements of a $^{14}$C-activated 
self-luminous phosphor source.
Instrument pointings (coordinates of FOV centers) are computed from the spacecraft attitude parameters.
The brightness contribution of bright stars is subtracted referring to stars in the Yale Bright Star Catalog
\citep{hoffleit64} and $V < 8$ stars in the U.S. Naval Observatory (USNO) Photoelectric Catalog \citep{blanco68}.
Finally, observed star crossings and in-flight measurements of the internal phosphor source are used to correct 
for temporal variation of the instrument sensitivity and to revise telescope pointings.
The final position accuracy of FOVs is 0$^{\circ}$.15 -- 0$^{\circ}$.40.
The processed data are stored in the Background Sky Tape \citep{weinberg81}, which is now available from NASA's 
National Space Science Data Center.

Using the {\it Pioneer 10} IPP data collected at the various heliocentric distances $R$, \citet{hanner74} find
that the ZL brightness at $R$ = 2.41 AU is less than 10 \% of that observed at $R$ = 1 AU, and that the ZL brightness is 
below the detectable level of the instrument when the spacecraft is at or beyond $R$ = 3.26 AU.
Hence, the data obtained at larger distances are most suitable for an analyses of diffuse radiations outside the ZL clouds.
In Table \ref{tab:obslog} we summarize the IPP measurements carried out at $R \ge 3.26$ AU.

\begin{table*}
\begin{center}
\caption{Observation Journal of the {\it Pioneer 10/11} IPPs at Outside the ZL Clouds \label{tab:obslog}}
\begin{tabular}{ccccccc}
\tableline\tableline
           &           &  $R$\tablenotemark{a} & Spin Axis\tablenotemark{b} &            & Number of & Usable   \\
Spacecraft & Day/Year  & (AU)                  & (R.A., Decl.)              & Look Angle & Exposures & Fraction  \\
\tableline
{\it Pioneer 10} & 354/1972 & 3.26 & (108$^{\circ}$.83, 23$^{\circ}$.62) &  30$^{\circ}$.75 -- 170$^{\circ}$.11 & 5,696 & 60.3 \% \\
{\it Pioneer 10} & 093/1973 & 3.92 & (138$^{\circ}$.50, 17$^{\circ}$.54) & 128$^{\circ}$.62 -- 169$^{\circ}$.98 & 1,344 & 75.4 \% \\
{\it Pioneer 10} & 149/1973 & 4.22 & (141$^{\circ}$.99, 16$^{\circ}$.68) &  28$^{\circ}$.63 -- 170$^{\circ}$.66 & 5,504 & 61.4 \% \\
{\it Pioneer 10} & 216/1973 & 4.54 & (135$^{\circ}$.03, 19$^{\circ}$.70) &  91$^{\circ}$.66 -- 170$^{\circ}$.07 & 2,816 & 50.2 \% \\
{\it Pioneer 10} & 237/1973 & 4.64 & (130$^{\circ}$.60, 19$^{\circ}$.71) &  29$^{\circ}$.00 -- 169$^{\circ}$.40 & 5,312 & 56.6 \% \\
{\it Pioneer 10} & 279/1973 & 4.81 & (127$^{\circ}$.40, 20$^{\circ}$.25) &  28$^{\circ}$.92 -- 169$^{\circ}$.22 & 5,248 & 57.4 \% \\
{\it Pioneer 10} & 021/1974 & 5.08 & (143$^{\circ}$.97, 15$^{\circ}$.21) &  30$^{\circ}$.64 -- 169$^{\circ}$.51 & 5,504 & 58.4 \% \\
{\it Pioneer 11} & 057/1974 & 3.50 & (155$^{\circ}$.13, 11$^{\circ}$.95) &  29$^{\circ}$.65 -- 169$^{\circ}$.00 & 4,544 & 72.6 \% \\
{\it Pioneer 10} & 068/1974 & 5.15 & (155$^{\circ}$.88, 10$^{\circ}$.23) &  28$^{\circ}$.92 -- 169$^{\circ}$.87 & 5,376 & 67.2 \% \\
{\it Pioneer 11} & 106/1974 & 3.81 & (166$^{\circ}$.08,  7$^{\circ}$.67) &  29$^{\circ}$.63 -- 170$^{\circ}$.64 & 4,672 & 73.5 \% \\
{\it Pioneer 11} & 148/1974 & 4.06 & (173$^{\circ}$.70,  4$^{\circ}$.63) &  29$^{\circ}$.56 -- 170$^{\circ}$.65 & 4,672 & 77.1 \% \\
{\it Pioneer 11} & 178/1974 & 4.22 & (175$^{\circ}$.11,  4$^{\circ}$.12) &  29$^{\circ}$.54 -- 170$^{\circ}$.55 & 4,608 & 72.9 \% \\
{\it Pioneer 11} & 236/1974 & 4.51 & (171$^{\circ}$.01,  5$^{\circ}$.94) &  29$^{\circ}$.61 -- 169$^{\circ}$.93 & 4,544 & 80.8 \% \\
{\it Pioneer 11} & 267/1974 & 4.66 & (165$^{\circ}$.58,  8$^{\circ}$.00) &  31$^{\circ}$.45 -- 168$^{\circ}$.39 & 4,416 & 70.4 \% \\
\tableline
\end{tabular}
%% Any table notes must follow the \end{tabular} command.
\tablecomments{$^{\rm a}$Heliocentric distance of the spacecraft.
  $^{\rm b}$The end of the spin axis pointing toward the Earth.}
%\tablenotetext{a}{Heliocentric distance of the spacecraft.}
%\tablenotetext{b}{The end of the spin axis pointing toward the earth.}
%\tablecomments{See text for the explanation of the flags.}
\end{center}
\end{table*}

\subsection{Data Assessment}

\begin{figure}
\epsscale{1.0}
\plotone{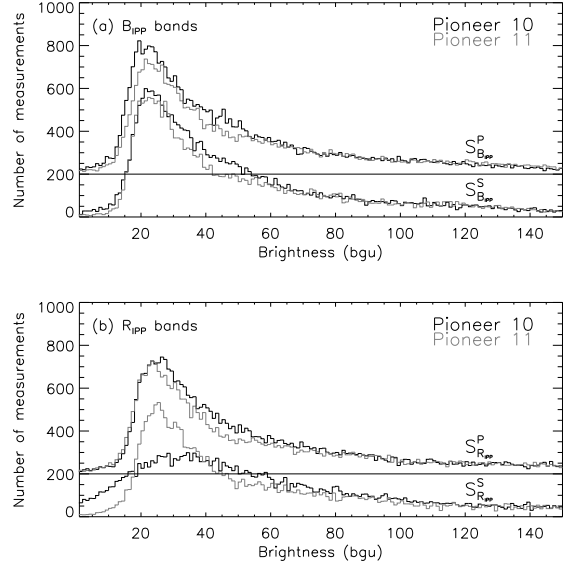}%{../../IPPdata/summary_p1.ps}
\caption{Distributions of the $S^{\rm S}_{B_{\rm IPP}}$, $S^{\rm P}_{B_{\rm IPP}}$ (panel (a)) and 
  $S^{\rm S}_{R_{\rm IPP}}$, $S^{\rm P}_{R_{\rm IPP}}$ (panel (b)) brightness for the P10 (black lines) and P11 
  (gray lines) measurements.
  Those of the $S^{\rm P}_{B_{\rm IPP}}$ and $S^{\rm P}_{R_{\rm IPP}}$ are shifted upward by $+$200 for visibility.
  \label{fluxdistr}}
\end{figure}

First, we look through all the measurements in Table \ref{tab:obslog} in order to identify and remove the poor-quality data.
%assign each of them a quality flag.
%Initially, all the measurements are given the flag 0 ("usable data").
Recall that one IPP exposure generates four measurements, i.e., $S^{\rm P}$ and $S^{\rm S}$ at each of $B_{\rm IPP}$ and $R_{\rm IPP}$.
We remove the exposures with a negative flux record in at least one of the four measurements, which are most likely the corrupted 
data due to interrupted data communication between the spacecrafts and the ground station.
Also removed are the exposures taken when the angle between the instrument pointing direction and the sun ("elongation") 
is less than 70$^{\circ}$ for the {\it Pioneer 10} and 45$^{\circ}$ for the {\it Pioneer 11}.
These exposures could be subject to the scattered sunlight incident on the instruments \citep{weinberg81}.
%As described above, 64 exposures (or "sectors") constitute one data roll.
We then calculate median and standard deviation of the four flux ratios,
$S^{\rm P}_{B_{\rm IPP}}$/$S^{\rm S}_{B_{\rm IPP}}$, $S^{\rm P}_{R_{\rm IPP}}$/$S^{\rm S}_{R_{\rm IPP}}$,
$S^{\rm P}_{B_{\rm IPP}}$/$S^{\rm P}_{R_{\rm IPP}}$, and $S^{\rm S}_{B_{\rm IPP}}$/$S^{\rm S}_{R_{\rm IPP}}$,
of 64 exposures constituting one data roll, and remove all the exposures in the data roll 
showing the outlying values. % in either of the four flux ratios.
%At this stage, we decide to discard all the P10 $R_{\rm IPP}$-band $S^{\rm S}$ data since they are found to be
%exceptionally noisy.
In Figure \ref{fluxdistr} we show the distributions of the $S^{\rm S}_{B_{\rm IPP}}$, $S^{\rm P}_{B_{\rm IPP}}$,
$S^{\rm S}_{R_{\rm IPP}}$, and $S^{\rm P}_{R_{\rm IPP}}$ brightness for the P10 and P11 measurements after the above removal of
the poor-quality data.
Overall they show the similar distributions to each other, while that of the P10 $S^{\rm S}_{R_{\rm IPP}}$ is significantly different 
from others.
It is due to the exceptionally large noise in this channel, therefore we decide to discard all the P10  $S^{\rm S}_{R_{\rm IPP}}$ data
at this stage.

\begin{figure}
\epsscale{1.0}
\plotone{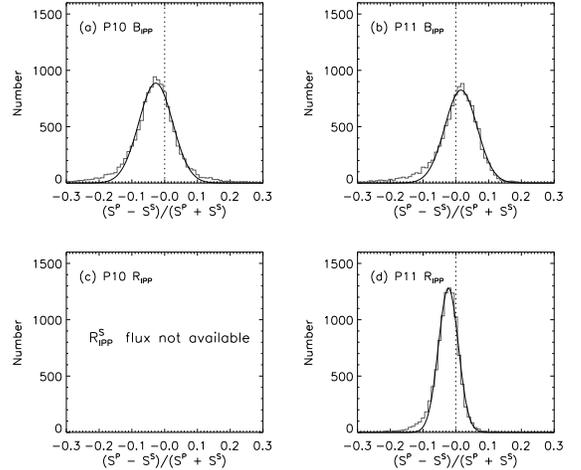}%{../IPPdata/summary_p3.ps}
\caption{Distributions of the brightness deviation $D$ = $(S^{\rm P} - S^{\rm S})$/$(S^{\rm P} + S^{\rm S})$ in 
  the P10 $B_{\rm IPP}$ (panel (a)), P11 $B_{\rm IPP}$ (panel (b)), and P11 $R_{\rm IPP}$ (panel (d)) measurements (histograms). 
  The dotted lines are drawn at $D$ = 0.
  The superposed smooth curves are the Gaussian distributions fitted to the data.\label{fluxcompare}}
\end{figure}

The dominant contributor to measured IPP brightness is Galactic stars.
Since integrated starlight over an FOV of $\sim$10 deg$^2$ is expected to polarize very little on average, the two orthogonal 
polarization components $S^{\rm P}$ and $S^{\rm S}$ have essentially the equal brightness.
Therefore the comparison between the two, which were measured by the different detector units at the end of the different optical 
paths, can be used to estimate accuracy of the IPP measurements.
We plot in Figure \ref{fluxcompare} distributions of the brightness deviation
\begin{eqnarray*}
D = \frac{S^{\rm P} - S^{\rm S}}{S^{\rm P} + S^{\rm S}} 
= \frac{S^{\rm P} - S/2}{S/2}
= - \frac{S^{\rm S} - S/2}{S/2} ,
\end{eqnarray*}
where $S = S^{\rm P} + S^{\rm S}$ is the total sky brightness at $B_{\rm IPP}$ or $R_{\rm IPP}$, in the brightness range 
$S$ = 30 -- 90 bgu where this work focuses on.
Narrowing the range to $S$ = 30 -- 50, 50 -- 70, or 70 -- 90 bgu does not change the following estimates significantly.
Figure \ref{fluxcompare} shows that the distributions of $D$ are close to the Gaussian distribution,
% with the best-fit mean and standard deviation of ($D_0$, $\sigma_D$) = ($-$0.027, 0.052) for P10 $B_{\rm IPP}$, (0.016, 0.049) 
%for P11 $B_{\rm IPP}$, and ($-$0.023, 0.030) for P11 $R_{\rm IPP}$.
therefore mean and standard deviation of the best-fit Gaussian functions give an appropriate estimate of systematic and 
random uncertainties inherent in the IPP measurements.
Thus, obtained values of the fractional systematic and random errors are 3\% and 6\% at $B_{\rm IPP}$ and 
3\% and 4\% at $R_{\rm IPP}$, respectively, relative to $S^{\rm P}$ or $S^{\rm S}$.
These values are divided by 2$\sqrt{2}$\footnote{
The factor 2 accounts for the change to the relative values to $S$, from relative to $S^{\rm P}$ or $S^{\rm S}$, while 
the factor $\sqrt{2}$ comes from the reduction of errors as the independent measurements $S^{\rm P}$ and $S^{\rm S}$ are summed.}
to give the corresponding errors relative to the total sky brightness $S = S^{\rm P} + S^{\rm S}$.
The derived estimates are summarized in Table \ref{tab:error} (first row) along with other estimates derived below.

\begin{table*}
\begin{center}
\caption{Error Estimates of the IPP Measurements \label{tab:error}}
\begin{tabular}{lcccc}
\tableline\tableline
%                   & $B_{\rm IPP}$  & $B_{\rm IPP}$ & $R_{\rm IPP}$ & $R_{\rm IPP}$  \\
                   & \multicolumn{2}{c}{$B_{\rm IPP}$} & \multicolumn{2}{c}{$R_{\rm IPP}$}  \\
Subset Criteria    & Systematic     & Random        & Systematic    & Random        \\
\tableline
Polarization Component  & 0.01       & 0.02          & 0.01          & 0.01          \\
Spacecraft              & 0.01       & 0.02          & 0.001         & 0.02          \\
Heliocentric Distance   & 0.008      & 0.02          & 0.0001        & 0.02          \\
Elongation              & 0.003      & 0.02          & 0.008         & 0.02          \\
Look Angle              & 0.001      & 0.02          & 0.003         & 0.02          \\
Sector                  & 0.01       & 0.02          & 0.005         & 0.02          \\
\tableline
\end{tabular}
%% Any table notes must follow the \end{tabular} command.
%\tablenotetext{a}{Heliocentric distance of the spacecraft.}
\tablecomments{Given are fractional errors relative to the total sky brightness $S = S^{\rm P} + S^{\rm S}$.}
\end{center}
\end{table*}

As already introduced, the total sky brightness is the sum of the two polarization components: $S = S^{\rm P} + S^{\rm S}$.
We substitute the P10 $S^{\rm P}_{R_{\rm IPP}}$ fluxes for the discarded P10 $S^{\rm S}_{R_{\rm IPP}}$ fluxes, i.e., 
$S_{R_{\rm IPP}} = 2 S^{\rm P}_{R_{\rm IPP}}$ for the P10 measurements.
We group each exposure with the neighbors within 5$^{\circ}$, and reject it if it deviates more than three times the standard deviation 
from the median of the group either in $S_{B_{\rm IPP}}$, $S_{R_{\rm IPP}}$, or $S_{B_{\rm IPP}}$/$S_{R_{\rm IPP}}$.
This process reveals that the P10 measurements on the day 216/1973 are very noisy, with nearly 50 \% rejected, 
in contrast to at most $\sim$10\% rejected for other days.
Therefore all the data taken on this day are discarded.
Table \ref{tab:obslog} lists fractions of the usable exposures as well as the total number of exposures taken on each day.
We find that approximately 70\% of the IPP data are usable for the present purpose.

\subsection{Mapping Process \label{sec:mapgeneration}}

We integrate all the usable data into a single sky map at each band.
Since IPP FOVs follow complex trajectories on the celestial sphere and most of the sky were observed several times or more, 
the final maps have much higher sensitivity and spatial resolution than the individual measurements.
The map generation is limited to the high Galactic latitude sky ($|b| > 35^{\circ}$) where foreground emissions from Galactic 
sources are relatively faint.

%We use only the flag 0 data for the map creation.
First, a set of pixels is placed on the celestial sphere based on the Galactic coordinate system, using Lambert's 
zenithal equal-area projection \citep[e.g.,][]{calabretta02}.
It is the same projection as used by \citet{sfd98} for creating the 100 $\mu$m emission map described below.
Pixel size of 0$^{\circ}$.32 $\times$ 0$^{\circ}$.32 is adopted, which projects all-sky data onto two 512 $\times$ 512 pixel 
maps centered on the north and south Galactic poles.
%(note that the two maps have some overlap with each other).
The brightness $S_{i,j}$ at each pixel ($i$, $j$) is calculated by
\begin{equation}
  S_{i,j} = \langle S_{i,j}^n \rangle ,
  \label{eq:wholemap}
\end{equation}
where $S_{i,j}^n$ represents an ensemble of the $n = 1, 2, ..., N$th IPP measurements whose FOVs contain the pixel ($i$, $j$)
and $\langle S_{i,j}^n \rangle$ is mean brightness of the ensemble.
When calculating $\langle S_{i,j}^n \rangle$, we apply the following rejection criteria based on diffuse emission 
brightness $S^{\rm diffuse} = S - S^{\rm star}$, where $S^{\rm star}$ is starlight brightness in each FOV
calculated in Section \ref{sec:starlight}\footnote{
It is better to use $S^{\rm diffuse}$ for the rejection process than to use $S$ itself because the latter can deviate 
significantly from each other in the ensemble $S_{i,j}^n$ due to different bright stars falling on the FOVs.
}.
We reject the data deviating more than three times the standard deviation from the mean of the ensemble either 
in $S^{\rm diffuse}_{B_{\rm IPP}}$,  $S^{\rm diffuse}_{R_{\rm IPP}}$, or 
$S^{\rm diffuse}_{B_{\rm IPP}}$/$S^{\rm diffuse}_{R_{\rm IPP}}$.
The unreliable measurements with $S < S^{\rm star} - 20$ bgu are also rejected
%whose $S$ value is smaller than $S^{\rm star}$ by more than 20 bgu 
(see Section \ref{sec:starlight}).
Associated error at each pixel is estimated from the scatter of the ensemble $S_{i,j}^n$.
This mapping algorithm is similar to first iteration of that adopted by \citet{gordon98}, and to
first iteration of the "maximum correlation method" \citep{aumann90} used for improving angular resolution of
the {\it Infrared Astronomy Satellite} ({\it IRAS}) maps.
Considering relatively low signal-to-noise ratios at low brightness regions where our study focuses on, 
we do not perform the higher-order iteration of the map generation algorithm in the present analysis.

\begin{figure*}
\epsscale{1.0}
\plotone{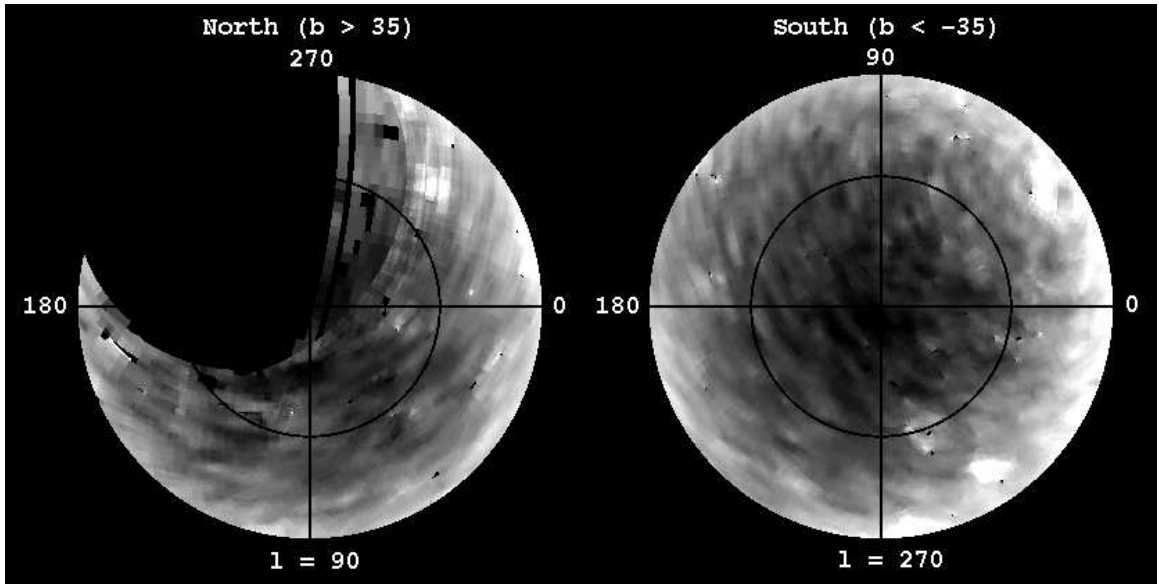}%{map_fig1.eps}
\caption{{\it Pioneer 10/11} IPP $B_{\rm IPP}$-band maps of the north (left) and south (right) 
  high Galactic latitude sky ($|b| > 35^{\circ}$).
  Color is linearly scaled between 30 (black) and 75 (white) bgu.
  The black lines show the Galactic longitudes of $l$ = 0$^{\circ}$, 90$^{\circ}$, 180$^{\circ}$, and 270$^{\circ}$,
  while the black circles show the Galactic latitudes of $b$ = $\pm$60$^{\circ}$.
  The Galactic poles are at the centers of the two images.\label{bluemap}}
\end{figure*}

\begin{figure*}
\epsscale{1.0}
\plotone{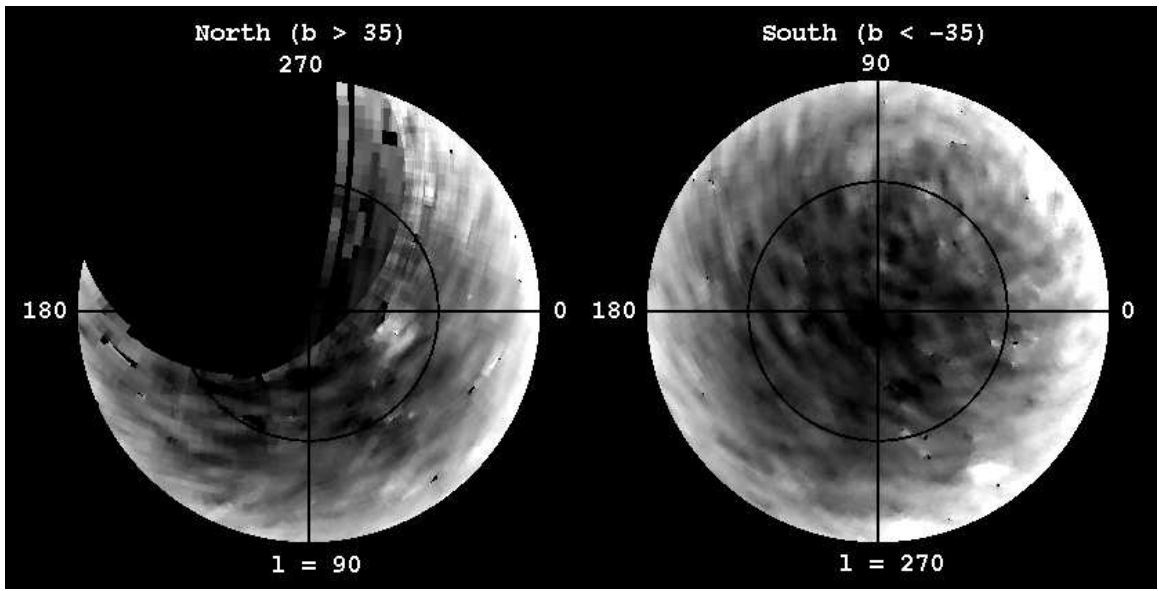}%{map_fig2.eps}
\caption{Same as Figure \ref{bluemap}, but for the IPP $R_{\rm IPP}$ band.
  Color is linearly scaled between 35 (black) and 80 (white) bgu.\label{redmap}}
\end{figure*}

We show the created $B_{\rm IPP}$- and $R_{\rm IPP}$-band maps of the high Galactic latitude sky ($|b| > 35^{\circ}$)
in Figures \ref{bluemap} and \ref{redmap}.
The above process projects $\sim$40,000 IPP exposures onto $\sim$200,000 pixels ($\sim$20,000 deg$^2$) at 
the high Galactic latitudes, which gives the rough estimate of the resultant angular resolution $\sim 0^{\circ}.7$.
%, hence the binning of at least 5 pixels 
%should be applied to the map in order to ensure the independency of the resultant IPP data points (though the number of 
%contributed IPP exposures per pixel is not perfectly uniform throughout the map).
On average, 16 measurements contribute to each pixel except in the fields with no observations.
Finally we apply 4 $\times$ 4 pixel binning to the created maps.

\begin{figure}
\epsscale{1.0}
\plotone{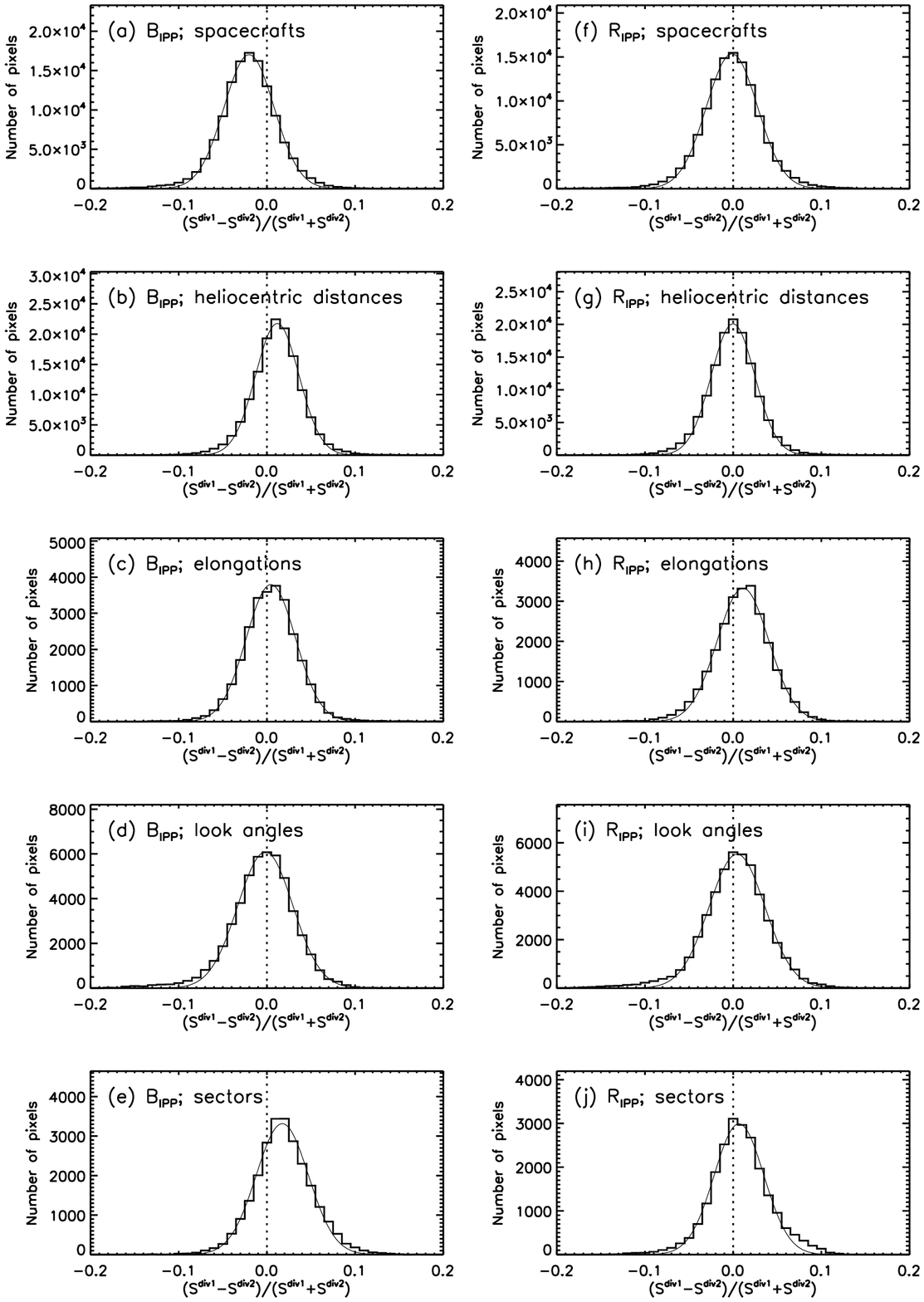}%{../map_div.ps}
\caption{Distributions of the brightness deviation $D$ = $(S^{\rm div1} - S^{\rm div2})$/$(S^{\rm div1} + S^{\rm div2})$ at
  the $B_{\rm IPP}$ (panels (a)--(e)) and $R_{\rm IPP}$- (panels (f)--(j)) bands (histograms).
  Division of the data are based on spacecrafts (panels (a) and (f)), the heliocentric distances (panels (b) and (g)),
  elongations (panels (c) and (h)), look angles (panels (d) and (i)), and sectors (panels (e) and (j)).
  See the text for the full description.
  The dotted lines are drawn at $D$ = 0.
  The superposed smooth curves are the Gaussian distributions fitted to the data.\label{map_div}}
\end{figure}

At this stage we can further check accuracy of the IPP measurements by investigating reproducibility of the created maps.
We divide the data into two subsets based on each of the following five criteria, which are selected to {\it maximize} 
systematic difference between the subsets; spacecrafts (P10 or P11); the heliocentric distances $R$
($< 4.5$ AU or $> 4.5$ AU); elongations ($< 115^{\circ}$ or $> 115^{\circ}$); look angles
($|L - 90^{\circ}| < 30^{\circ}$ or $> 30^{\circ}$); and sectors (01 -- 32 or 33 -- 64).
The demarcations of the heliocentric distances (4.5 AU), elongations ($115^{\circ}$), and look angles ($30^{\circ}$) 
are chosen so that the divided two subsets have similar numbers of measurements.
Then a pair of sky maps are created adopting each of the five divisions, and distributions of the brightness deviation
\begin{eqnarray*}
D = \frac{S^{\rm div1} - S^{\rm div2}}{S^{\rm div1} + S^{\rm div2}} 
= \frac{S^{\rm div1} - \bar{S}}{\bar{S}}
= - \frac{S^{\rm div2} - \bar{S}}{\bar{S}}
\end{eqnarray*}
are calculated in the overlapping regions.
Here, $S^{\rm div1}$ and $S^{\rm div2}$ are brightness of same pixels of the two maps and $\bar{S} = (S^{\rm div1} + S^{\rm div2})/2$.
The comparison is limited to the pixels for which more than three measurements contribute and within the brightness range
$\bar{S}$ = 30 -- 90 bgu.
The results are shown in Figure \ref{map_div}.
All the distributions are close to the Gaussian distribution; the resultant estimates of systematic and random errors from 
the best-fit mean and standard deviation values are summarized in Table \ref{tab:error}.
The table shows that any subset of the data reproduces the final sky map within 1 \% systematic uncertainty in both of 
$B_{\rm IPP}$ and $R_{\rm IPP}$.
Thus, we conclude that systematic errors inherent in the IPP measurements are within 1 \% of the total sky brightness.

\section{Galactic Light \label{sec:gallight}}

\subsection{Starlight \label{sec:starlight}}

Outside the detectable ZL clouds, the dominant brightness component incident on the IPPs is Galactic starlight.
Contribution of bright stars has already been subtracted when the Background Sky Tape was created, referring to 
stars in the Yale Bright Star Catalog and $V < 8$ stars in the USNO Photoelectric Catalog (see Section \ref{sec:IPPobs}).
Therefore, we have to subtract contribution of fainter stars from the IPP measurements.
For this purpose, we use two all-sky star catalogs along with a star-count model as described below.
Transformations of stellar colors between the different filter systems are also developed.

Integrated brightness of relatively bright ($V \la 11$ mag) stars are calculated using the {\it Tycho-2} Catalog \citep{hog00a}.
The catalog is based on observations by the star mapper of the ESA {\it Hipparcos} satellite, and contains 
positions, proper motions, and photometry data for $\sim$2,500,000 brightest stars in the sky.
Source magnitudes were measured at the two passbands $B_T$ and $V_T$, which are close to the Johnson $B$ and $V$ bands.
Detection completeness is $\sim$90 \% at $V = 11.5$ mag, with magnitude errors less than 0.1 mag for brighter stars.
Systematic uncertainty of the photometry is estimated to be smaller than 0.02 mag \citep{hog00b}.

Contribution of fainter ($V \ga 11$ mag) stars is derived from the {\it HST} Guide Star Catalog II (GSC-II) version 2.3 
\citep{lasker08}.
It was constructed from the scanned images of 9541 Palomar and UK Schmidt photographic sky survey plates
digitized at Space Telescope Science Institute (STScI).
About 950,000,000 sources to the limiting magnitude of $R \sim 20$ mag are contained in the catalog with
photometry information mainly at the three natural plate passbands $B_J$, $R_F$, and $I_N$.
The limiting magnitudes of the catalog correspond to $B_{\rm IPP}$ = 22.5 mag and $R_{\rm IPP}$ = 20.5 mag (see below for
transformations of stellar magnitudes).
Magnitude errors are less than 0.2 mag at $R_F < 19.5$ mag, and 
systematic uncertainty of the photometry is estimated to be within a few hundredths of magnitude.

We estimate contribution of stars even fainter than the GSC-II detection limits using a star-count model provided 
by a stellar population synthesis code TRILEGAL \citep{girardi05}.
TRILEGAL generates and spatially distributes stars according to the user-defined parameters, such as star formation rate, 
age--metallicity relation, initial mass function, and distribution geometry, and then predicts stellar photometry at
any field of the Galaxy (also at star clusters and nearby galaxies).
We adopt the default set of parameters for deriving Galactic star counts at the Bessel $B$ and $R$ bands down to 32.0 mag.
Then the ratios $S^{\rm faint}$/$S^{\rm bright}$ are calculated at 30 representative Galactic coordinates, where 
$S^{\rm bright}$ and  $S^{\rm faint}$ represent integrated brightness of stars in the brightness ranges 
$B_{\rm IPP}$ ($R_{\rm IPP}$) = 15.0 -- 22.5 (15.0 -- 20.5) mag and 22.5 -- 32.0 (20.5 -- 32.0) mag, respectively.
Note that the demarcations between bright and faint stars correspond to the GSC-II limiting magnitudes.
We find that these faint star fractions are small, ranging from $S^{\rm faint}$/$S^{\rm bright}$ = 0.01 to 0.03 at $B_{\rm IPP}$
and from $S^{\rm faint}$/$S^{\rm bright}$ = 0.03 to 0.12 at $R_{\rm IPP}$.
The ratios vary fairly smoothly between the 30 positions, hence we calculate those at any positions
by interpolating among the three neighbors chosen from the 30 points.
The effects of the uncertainty in the above process for estimating contribution of the faintest stars on our final
results are discussed in Section 5.

%Groenewegen et al. A&A, 2002, 392, 741 for simple description of the code...
%http://www.aanda.org/index.php?option=com_article&access=bibcode&Itemid=129&bibcode=2002A%2526A...392..741GFUL

\begin{figure*}
\epsscale{1.0}
\plotone{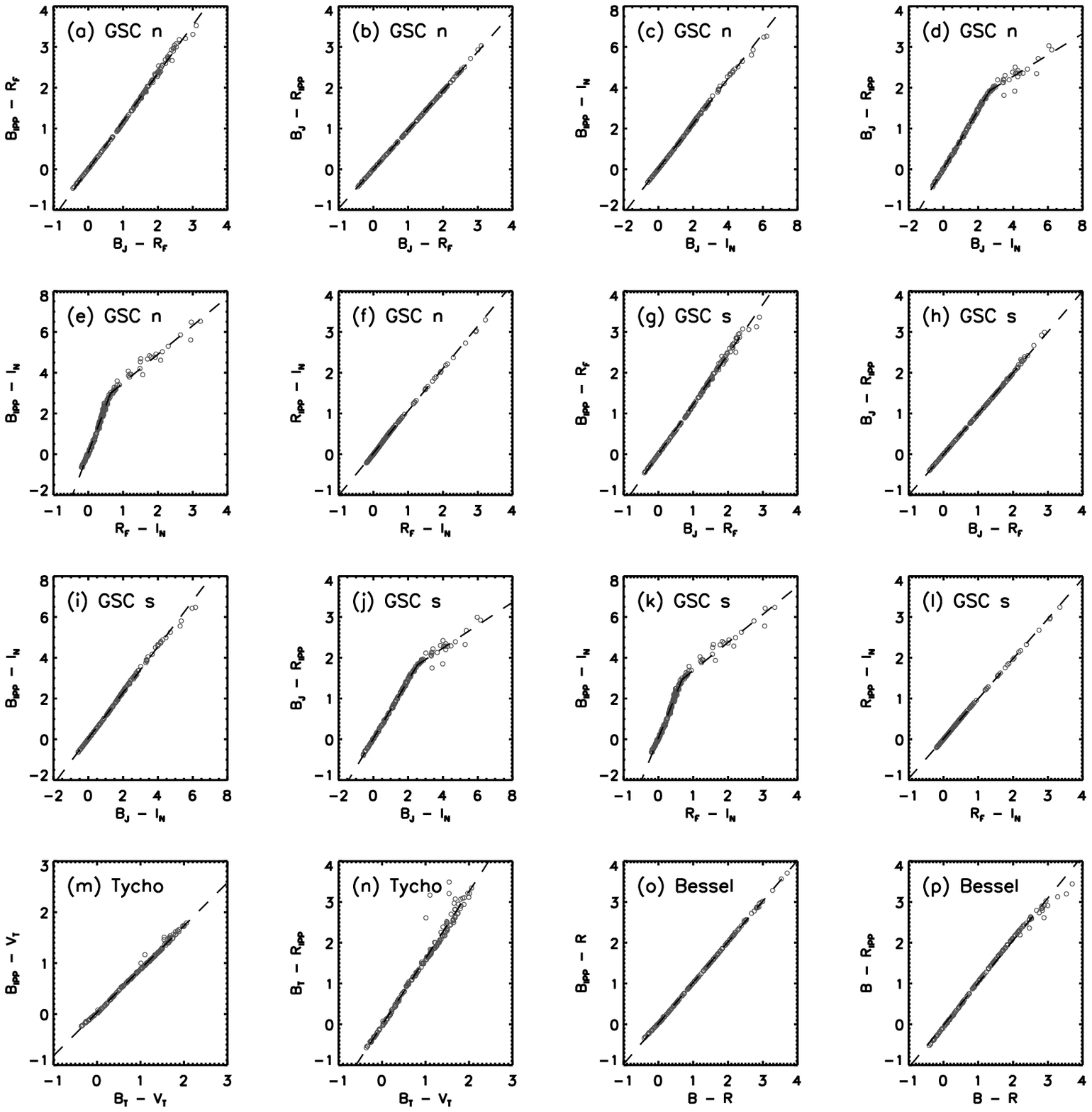}%{colorconv_fig.ps}
\caption{Conversions of stellar colors between the {\it Pioneer} IPP and the star catalog systems:
  the Palomar natural plate passbands for the GSC-II northern stars (panels (a)--(f)), 
  the UK Schmidt natural plate passbands for the GSC-II southern stars (panels (g)--(l)), 
  the {\it Hipparcos} passbands for the {\it Tycho-2} stars (panels (m) and (n)),
  and the Bessel passbands (panels (o) and (p)).
  The open circles represent the 175 BPGS stars while the dashed lines show the regression lines.\label{colorconv}}
\end{figure*}

Transformations of stellar magnitudes from the {\it Hipparcos} ($B_T$ and $V_T$), the natural plate band ($B_J$, $R_F$, 
and $I_N$), and the Bessel ($B$ and $R$) systems to the IPP ($B_{\rm IPP}$ and $R_{\rm IPP}$) system are developed using 
the Bruzual-Persson-Gunn-Stryker (BPGS) stellar spectral atlas provided by STScI.
The atlas contains the optical-to-near-IR spectra of 175 stars with a wide range of spectral type and luminosity class,
and is often used to simulate color--color diagrams of various Galactic stars \citep[e.g.,][]{hewett06}.
Figure \ref{colorconv} shows the relations of stellar colors in the different passband systems.
All the relations can be well described by one- or two-component regression lines, which are used for 
converting stellar colors in this work.
For convenience, we summarize the coefficients of the regression lines in Table \ref{tab:colorconv}.

\begin{table*}
\begin{center}
\caption{Conversion of Stellar Colors \label{tab:colorconv}}
\begin{tabular}{cccccc}
\tableline\tableline
Color 1 (To)          & Color 0 (From)  & $a$  & $b$  &  & Filter System (From)\\
\tableline
$B_{\rm IPP} - R_F$  & $B_J - R_F$ & 0.00    & 1.18 & & GSC-II north \\
$B_J - R_{\rm IPP}$  & $B_J - R_F$ & 0.00    & 0.96 & & GSC-II north  \\
$B_{\rm IPP} - I_N$  & $B_J - I_N$ & 0.03    & 1.11 & & GSC-II north \\
$B_J - R_{\rm IPP}$  & $B_J - I_N$ & 0.01    & 0.72 & at [color 0] $<$ 2.66 & GSC-II north\\
                     &             & 1.23    & 0.26 & at [color 0] $>$ 2.66 & GSC-II north \\
$B_{\rm IPP} - I_N$  & $R_F - I_N$ & 0.07    & 4.56 & at [color 0] $<$ 0.64 & GSC-II north\\
                     &             & 2.09    & 1.41 & at [color 0] $>$ 0.64 & GSC-II north \\
$R_{\rm IPP} - I_N$  & $R_F - I_N$ & 0.03    & 1.04 & & GSC-II north\\
\tableline
$B_{\rm IPP} - R_F$  & $B_J - R_F$ & 0.00    & 1.23 & & GSC-II south  \\
$B_J - R_{\rm IPP}$  & $B_J - R_F$ & $-$0.01 & 1.00 & & GSC-II south  \\
$B_{\rm IPP} - I_N$  & $B_J - I_N$ & 0.04    & 1.13 & & GSC-II south  \\
$B_J - R_{\rm IPP}$  & $B_J - I_N$ & 0.02    & 0.71 & at [color 0] $<$ 2.54 & GSC-II south\\
                     &             & 1.11    & 0.28 & at [color 0] $>$ 2.54 & GSC-II south\\
$B_{\rm IPP} - I_N$  & $R_F - I_N$ & 0.05    & 4.23 & at [color 0] $<$ 0.68 & GSC-II south \\
                     &             & 2.00    & 1.38 & at [color 0] $>$ 0.68 & GSC-II south \\
$R_{\rm IPP} - I_N$  & $R_F - I_N$ & 0.01    & 0.98 & & GSC-II south\\
\tableline
$B_{\rm IPP} - V_T$  & $B_T - V_T$ & 0.03    & 0.85 & & {\it Tycho-2}  \\
$B_T - R_{\rm IPP}$  & $B_T - V_T$ & $-$0.02 & 1.64 & & {\it Tycho-2}  \\
\tableline
$B_{\rm IPP} - B_R$  & $B - R$     & 0.02    & 1.00 & & Bessel \\
$B_B - R_{\rm IPP}$  & $B - R$     & $-$0.03 & 1.04 & & Bessel  \\
\tableline
\end{tabular}
%% Any table notes must follow the \end{tabular} command.
%\tablenotetext{a}{Heliocentric distance of the spacecraft.}
\tablecomments{Conversion from the color 0 to the color 1: [color 1] = $a$ + $b$ $\times$ [color 0].}
\end{center}
\end{table*}

Using the above star catalogs and star-count model with the stellar color conversions, we compute integrated brightness 
of starlight $S^{\rm star}$ falling on FOV of each IPP measurement. % in the $B_{\rm IPP}$ and $R_{\rm IPP}$ bands.
In order to avoid duplicated counting, all the stars found in the Yale Bright Star Catalog and the $V < 8$ stars
in the USNO Photoelectric Catalog are excluded from the {\it Tycho-2} and GSC-II catalogs.
The {\it Tycho-2} stars in the GSC-II are also removed.
Fractions of the fainter stars than the GSC-II detection limits derived from the TRILEGAL code, 
$S^{\rm faint}$/$S^{\rm bright}$, are used in combination with the actual values of $S^{\rm bright}$ to estimate 
contributions of the faintest stars down to 32.0 mag.
Then the starlight brightness is subtracted from each IPP FOV, resulting in diffuse emission brightness
$S^{\rm diffuse} = S - S^{\rm star}$.
Random uncertainties in the above process of starlight estimation and subtraction, including those originating from
the relatively large IPP pointing uncertainty, are directly reflected in the final data scatter as we see in Section 4.

At this stage, we find that a small fraction ($\sim$5 \%) of the IPP brightness are significantly {\it lower} than 
the starlight brightness of the fields, i.e., $S < S^{\rm star} -20$ bgu, which are roughly more than three times the standard 
deviation of the data scatter.
Since there are few data showing the opposite deviation $S > S^{\rm star} + 20$ bgu, it is most likely caused by some 
failures in the IPP measurements rather than the natural data scatter.
These data are rejected in the map generation process.

The diffuse emission brightness of IPP FOVs are integrated into sky maps using the same algorithm as described in 
Section \ref{sec:mapgeneration}, i.e.,
\begin{equation}
  S^{\rm diffuse}_{i,j}= \langle S_{i,j}^{{\rm diffuse}; n} \rangle ,
  \label{eq:diffusemap}
\end{equation}
where $S_{i,j}^{{\rm diffuse}; n}$ is an ensemble of the $n = 1, 2, ..., N$th IPP diffuse emission brightness whose FOVs contain the 
pixel ($i$, $j$).
The same rejection criteria as used in Equation (\ref{eq:wholemap}) are applied, and associated error at each pixel is estimated from 
the scatter of the ensemble $S_{i,j}^{{\rm diffuse}; n}$.
%Since the $S_n^*$'s are derived from the different $N$ sets of stars, the scatter in the diffuse emission brightness
%$S_n - S_n^*$ include the errors originating from the calculation of the starlight contributions.
Finally, 4 $\times$ 4 pixel binning is applied to the created diffuse emission maps.

\subsection{Diffuse Galactic Light \label{sec:dgl}}

\subsubsection{Overview}

The DGL is attributed to dust and gas particles in the interstellar medium (ISM).
Dominant contribution to the optical DGL comes from scattering of the interstellar radiation field (ISRF) by the dust.
This scattering process is the prime cause of the well-known interstellar extinction.
\citet{bernstein02a} and \citet{bernstein07} derive brightness of this scattered DGL component in their field
(Galactic coordinate $l = 206^{\circ}.6$ and $b = -59^{\circ}.8$) by combining a simple scattering model and 
thermal emission brightness at 100 $\mu$m wavelength.
The estimated value is $\sim 1$ bgu, which is a few tenths of the COB brightness claimed by them.

Another process generating the DGL is line and continuum emissions from warm ($T \sim 10^4$ K) ionized gas in the ISM.
The strongest line emission contributing to the optical DGL is H$\alpha$, which falls in our $R_{\rm IPP}$ band.
The fainter emission lines [\ion{N}{2}] $\lambda$6583 and [\ion{S}{2}] $\lambda$6716 could also contribute to the DGL, each with 
roughly $\sim$0.3 times the H$\alpha$ brightness \citep{reynolds85}.
Hydrogen Balmer lines other than H$\alpha$ (H$\beta$, H$\gamma$, ...) could contribute to the DGL in the $B_{\rm IPP}$ band.
Adopting the estimates of diffuse H$\alpha$ intensity provided by \citet{reynolds92}, the total DGL 
brightness due to the diffuse line emissions at $|b| = 60^{\circ}$ are less than 0.2 and 0.3 bgu at
$B_{\rm IPP}$ and $R_{\rm IPP}$, respectively (the case B condition of the recombination theory is assumed).
On the other hand, continuum emissions from warm ionized gas such as the two-photon, free--free, and bound--free emissions, 
are almost negligible.
\citet{bernstein02a} estimate the DGL brightness due to the continuum emissions as less than 0.01 bgu at the optical 
wavelengths longer than 0.37 $\mu$m.

%provides the rough estimate of the average H$\alpha$ brightness in the DGL as $2.9 \times 10^{-7}$
%cosec($|b|$) erg s$^{-1}$ cm$^{-2}$ sr$^{-1}$ at the Galactic latutudes $|b| \ge 15^{\circ}$.
%At $|b| > 35^{\circ}$, it corresponds to $< 0.5 \times 10^{-9}$ erg s$^{-1}$ cm$^{-2}$ \AA$^{-1}$ sr$^{-1}$ in the 
%$R_{\rm IPP}$ band.
%The fainter emission lines [N II] $\lambda$6583 and [S II] $\lambda$6716 also contribute to the eDGL in the $R_{\rm IPP}$ 
%band, both with roughly $\sim$0.3 times the brightness of H$\alpha$ \citep{reynolds85}.
%On the other hand, the hydrogen Balmer lines below H$\beta$ (H$\beta$, H$\gamma$, ...) make the major contribution to 
%the eDGL in the $B_{\rm IPP}$ band.
%Assuming the case B condition of the recombination theory and the H$\alpha$ brightness given above, their total
%$B_{\rm IPP}$-band brightness is $< 0.3 \times 10^{-9}$ erg s$^{-1}$ cm$^{-2}$ \AA$^{-1}$ sr$^{-1}$.
%(Ha:Hb:Hgama: ... = 2.85:1.00:0.469:0.260:0.159:0.105 is assumed!!!!!!).

In total, the DGL brightness at $B_{\rm IPP}$ and $R_{\rm IPP}$ are at most a few bgu at $|b| \sim 60^{\circ}$ 
and less at the higher Galactic latitudes, most of which comes from scattering of the ISRF by the interstellar dust.
%Since the DGL is caused by the interstellar dust and gas exposed to the ISRF, in principle one can eliminate its
%contribution by "observing" (directly or indirectly) the lines-of-sight (LOSs) free from these interstellar components.

We estimate the DGL contribution to IPP diffuse emission brightness by separating out the component which 
correlates with the diffuse Galactic far-IR emission.
It is well known that the far-IR emission traces thermal radiation from the interstellar dust heated by the ISRF.
Hence it is brighter in the fields with larger column density of the dust (and gas) and stronger ISRF, where 
the DGL is also brighter.
It naturally results in correlation between the DGL and far-IR brightness especially in optically thin dust regions,
where the both brightnesses are proportional to the dust column density.
Actually several authors find linear correlations between the DGL and the diffuse Galactic 100 $\mu$m brightness in different 
fields \citep{laureijs87,guhathakurta89,paley91}, while the reported DGL-to-100 $\mu$m brightness ratios vary 
from field to field.
As well as non-uniform dust properties, large measurement errors inherent in the treatments of the foreground emissions
are obviously one of the major contributors to this field-to-field variation.
With the DGL-to-100 $\mu$m brightness ratio $a_d$, the IPP diffuse emission brightness can be written as:
\begin{eqnarray}
  S^{\rm diffuse} &=&  S^{\rm DGL} + S^{\rm COB} \nonumber\\
  &=& a_d (S_{\rm 100{\mu}m}^{\rm diffuse} - S_{\rm 100{\mu}m}^{\rm CIB}) + S^{\rm COB} ,
  \label{eq:dgl}
\end{eqnarray}
where $S^{\rm DGL}$ and $S^{\rm COB}$ are the DGL and COB brightness at the IPP bands, and $S_{\rm 100{\mu}m}^{\rm diffuse}$ 
is diffuse 100 $\mu$m brightness observed outside the ZL clouds.
The cosmic infrared background (CIB) component $S_{\rm 100{\mu}m}^{\rm CIB}$ is subtracted from $S_{\rm 100{\mu}m}^{\rm diffuse}$ 
in the above equation, leaving only Galactic component.

\subsubsection{Simulation}

\begin{figure}
\epsscale{1.0}
\plotone{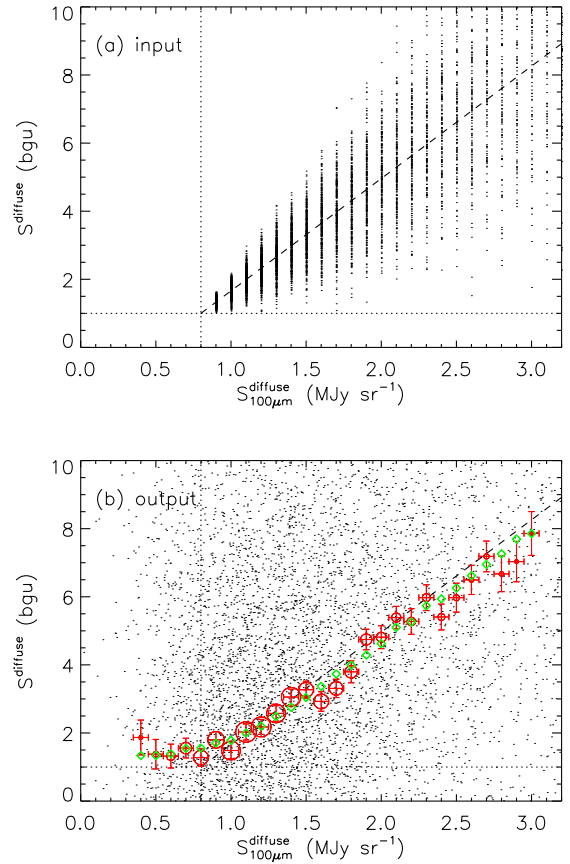}%{../vs100um_mcsimulate.ps}
\caption{Input (panel (a)) and output (panel (b)) $S^{\rm diffuse}$ and $S_{\rm 100{\mu}m}^{\rm diffuse}$ of one realization
  of the Monte Carlo simulation (dots).
  The dotted lines show the assumed COB and CIB brightness, while the dashed lines show the mean DGL-to-100$\mu$m relation 
  adopted (Equation (\ref{eq:dgl}) with $a_d = a_d^0$).
  The red circles and error bars represent mean values of the output $S^{\rm diffuse}$ and their errors in the
  $S_{\rm 100{\mu}m}^{\rm diffuse}$ bins of the realization, while the green diamonds represent mean values of all the 
  100 realizations.
  The sizes of the red circles are proportional to numbers of the data points in the $S_{\rm 100{\mu}m}^{\rm diffuse}$ bins.
  \label{vs100um_mc}}
\end{figure}

In order to make a proper interpretation of the observed relation between $S^{\rm diffuse}$ and $S_{\rm 100{\mu}m}^{\rm diffuse}$
in the presence of the data scatters, here we perform a simple Monte Carlo simulation.
We adopt the numbers and distributions as similar as possible to what is observed (found in Section 4), while
we do not intend to deduce any accurate numbers from this simulation.
The COB and CIB brightness are assumed to be $S^{\rm COB}$ = 1.0 bgu and $S_{\rm 100{\mu}m}^{\rm CIB}$ = 0.8 MJy sr$^{-1}$.
We generate $\sim$8000 mock data with 100 $\mu$m brightness following the simple probability distribution
$\propto$ exp$(-S_{\rm 100{\mu}m}^{\rm diffuse})$ at 
$S_{\rm 100{\mu}m}^{\rm diffuse} > S_{\rm 100{\mu}m}^{\rm CIB}$,
which results in the similar number distributions of $S_{\rm 100{\mu}m}^{\rm diffuse}$ to the observed at the high Galactic latitudes.
$S^{\rm diffuse}$ is calculated by Equation (\ref{eq:dgl}) assuming the Gaussian probability distribution of $a_d$ with the mean
$a_d^0$ and the standard deviation $\sigma_{a_d}$, where $a_d^0$ = 3.3 bgu (MJy sr$^{-1}$)$^{-1}$ and 
$\sigma_{a_d}$ = 1.0 bgu (MJy sr$^{-1}$)$^{-1}$ are adopted.
The former value is similar to what is found in the actual measurements while the latter is coupled with measurement errors and 
is hard to derive from the present data.
The resultant input relation between $S^{\rm diffuse}$ and $S_{\rm 100{\mu}m}^{\rm diffuse}$ is shown in the top panel of Figure 
\ref{vs100um_mc}.
Then the Gaussian random errors are added to $S^{\rm diffuse}$ and $S_{\rm 100{\mu}m}^{\rm diffuse}$ with the standard deviations of
5.0 bgu and 0.2 MJy sr$^{-1}$, respectively, which reproduces a similar $S^{\rm diffuse}$--$S_{\rm 100{\mu}m}^{\rm diffuse}$
plot to the observed.
We calculate mean values of $S^{\rm diffuse}$ and their errors in $S_{\rm 100{\mu}m}^{\rm diffuse}$ bins stepped by 
0.1 MJy sr$^{-1}$, which are shown in the bottom panel of Figure \ref{vs100um_mc} (red circles).
We repeat the simulation 100 times varying the random components, and obtain mean $S^{\rm diffuse}$--$S_{\rm 100{\mu}m}^{\rm diffuse}$ 
relation of the whole realizations as plotted in the same panel (green diamonds).

Figure \ref{vs100um_mc} shows that in the presence of the realistic data scatters (red circles), we will find the two distinct 
features in the observed $S^{\rm diffuse}$--$S_{\rm 100{\mu}m}^{\rm diffuse}$ plot.
At $S_{\rm 100{\mu}m}^{\rm diffuse} > S_{\rm 100{\mu}m}^{\rm CIB}$, the original correlation of the DGL component is simply
reproduced with some random deviations.
This correlation breaks at $S_{\rm 100{\mu}m}^{\rm diffuse} \simeq S_{\rm 100{\mu}m}^{\rm CIB}$, and a relatively flat 
distribution of $S^{\rm diffuse}$ is seen at the smaller $S_{\rm 100{\mu}m}^{\rm diffuse}$.
The $S^{\rm diffuse}$ value in this region is dominated by the COB brightness while a minor contribution of the DGL is present.
Hence, such a inflection point of the $S^{\rm diffuse}$--$S_{\rm 100{\mu}m}^{\rm diffuse}$ relation gives rough
estimates of the COB and CIB brightness.
Alternatively, if the CIB brightness $S_{\rm 100{\mu}m}^{\rm CIB}$ is known in advance, we can obtain more accurate estimates
of the COB brightness at the intersect of the regression line of the correlation at 
$S_{\rm 100{\mu}m}^{\rm diffuse} > S_{\rm 100{\mu}m}^{\rm CIB}$ and the line of 
$S_{\rm 100{\mu}m}^{\rm diffuse} = S_{\rm 100{\mu}m}^{\rm CIB}$.
Note that it corresponds to substituting $S_{\rm 100{\mu}m}^{\rm diffuse} = S_{\rm 100{\mu}m}^{\rm CIB}$ into Equation (\ref{eq:dgl}).

%While $a_{\rm d}$ varies from LOS to LOS, we show below that a single linear correlation between $S_{\rm 100{\mu}m}$ 
%and $S^{\rm DGL}$ is actually found when a sufficiently large number of LOSs are integrated.
%Eq. (\ref{eq:dgl}) points that the starting point of this correlation is offset from the coordinate origin on the
%$S_{\rm 100{\mu}m}$  versus $S_{\rm IPP}^{\rm diffuse}$ plot by the amounts of $S_{\rm 100{\mu}m}^{\rm CIB}$ and $S^{\rm COB}$.

%Then, we can subtract the DGL from the IPP measurements by finding the $S_{\rm IPP}$ value on this linear correlation where 
%$S^{\rm 100{\mu}m}$ is equal to $S^{\rm 100{\mu}m}_{\rm CIB}$ (i.e., no Galactic 100 $\mu$m emission; see Eq. \ref{eq:dgl}).

%Hence we can subtract the sDGL component by (i) finding out a linear correlation between the observed IPP brightness after
%the starlight subtraction ($S_{\rm IPP}$) and the 100 $\mu$m brightness ($S^{\rm 100{\mu}m}$), and (ii) deriving the IPP 
%brightness at where the 100 $\mu$m brightness is equal to the extragalactic contribution 
%($S_{\rm IPP} [S^{\rm 100{\mu}m} = S^{\rm 100{\mu}m}_{\rm CIB}]$).

\section{Results\label{sec:results}}

\begin{figure}
\epsscale{1.0}
\plotone{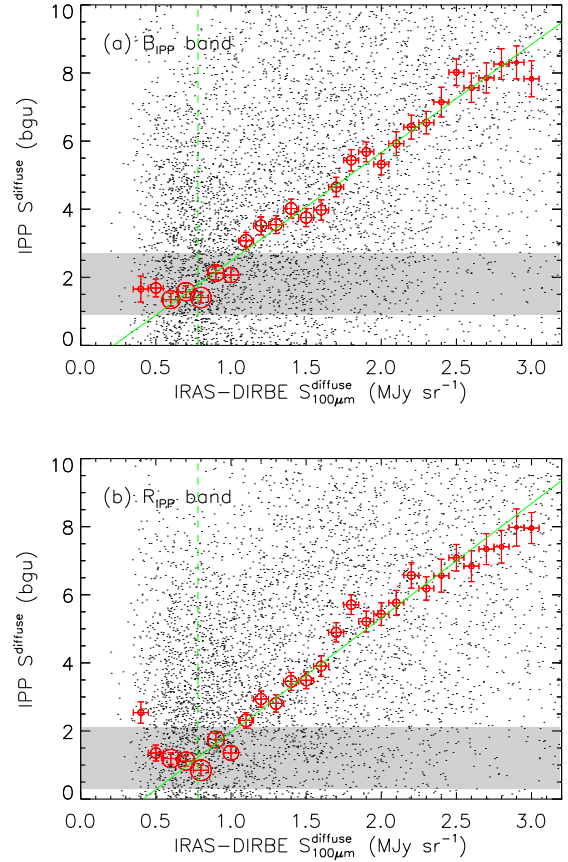}%{../vs100um_p1.ps}
\caption{Observed IPP diffuse emission brightness $S^{\rm diffuse}$ at $B_{\rm IPP}$ (panel (a)) and $R_{\rm IPP}$ 
  (panel (b)) vs. the diffuse 100 $\mu$m brightness $S_{\rm 100{\mu}m}^{\rm diffuse}$ (dots).
  The red circles and error bars represent mean values of $S^{\rm diffuse}$ and their errors in the $S_{\rm 100{\mu}m}^{\rm diffuse}$ bins.
  The sizes of the circles are proportional to the numbers of the data points in the bins.
  The solid green lines show the regression lines at $S_{\rm 100{\mu}m}^{\rm diffuse} >$ 1.0 MJy sr$^{-1}$, 
  while the dashed green lines show the CIB brightness $S_{\rm 100{\mu}m}^{\rm CIB}$ reported by \citet{lagache00}.
  The shaded areas show 1$\sigma$ confidence intervals of our final estimates of the COB brightness.\label{vs100um}}
\end{figure}

We show the measured IPP diffuse emission brightness $S^{\rm diffuse}$ versus the diffuse 100 $\mu$m brightness 
$S_{\rm 100{\mu}m}^{\rm diffuse}$ outside the ZL clouds in Figure \ref{vs100um}.
The present analysis focuses on the lowest brightness region with $S_{\rm 100{\mu}m}^{\rm diffuse}$ $<$ 3.0 MJy sr$^{-1}$
and with the IPP map coverage at the Galactic latitudes $|b| > 35^{\circ}$, which corresponds to about a quarter of 
the whole sky.
%which corresponds to about 80 \% of the high Galactic latitude ($|b| > 35^{\circ}$) sky.
Mean values of $S^{\rm diffuse}$ and their errors are calculated in $S_{\rm 100{\mu}m}^{\rm diffuse}$ bins stepped 
by 0.1 MJy sr$^{-1}$ (red circles), which are used for the regression analysis in what follows.
The errors are derived from the data scatters, which arise from the following three main causes: the IPP measurement errors, 
uncertainties in the estimation and the subtraction of Galactic starlight, and intrinsic scatters of DGL-to-100 $\mu$m 
brightness ratios.
The diffuse 100 $\mu$m brightness is taken from \citet{sfd98}.
They reprocess and combine the {\it IRAS} and the {\it Cosmic Background Explorer} 
({\it COBE}) Diffuse Infrared Background Experiment (DIRBE) measurements which results in an all-sky 100 $\mu$m brightness 
map with high angular resolution provided by the {\it IRAS} and accurate absolute flux calibration provided by the DIRBE.
The ZL and confirmed point sources and galaxies have been removed, while no significant CIB is detected in their analysis.
%Thus their 100 $\mu$m map can be regarded as a map of the infrared cirrus, with the possible residual contamination from
%faint Galactic stars and the CIB.
We apply 32 $\times$ 32 pixel binning to their 100 $\mu$m map so that it has the same resolution as our IPP maps.
Uncertainty of the 100 $\mu$m brightness is not explicitly given in \citet{sfd98}, while accuracy of the interstellar 
dust reddening ($E_{B-V}$) derived from the 100 $\mu$m map is estimated to be 16\%.
% based on the external check using the Mg$_2$ - $B - V$ color relation of elliptical galaxies.
Since $E_{B-V}$ has a larger uncertainty than the original 100 $\mu$m brightness due to the various assumptions made, 
e.g., those made in estimating dust temperatures, accuracy of the 100 $\mu$m brightness would be considerably better.
In addition, the 32 $\times$ 32 pixel binning applied in the present work should further reduce random errors.

While the better 100 $\mu$m brightness map is presented by \citet{miville05}, their Improved Reprocessing of the
{\it IRAS} Survey (IRIS) map includes point sources and thus cannot be compared directly to our diffuse emission maps.
However, a very good agreement is found between the IRIS and \citet{sfd98} data sets at scales larger than $\sim 0^{\circ}.5$ 
\citep{miville05} where the present analysis focuses on (note the IPP map resolution of $\sim 0^{\circ}.7$).
Hence, we do not expect a significant change of the final results when the IRIS map is alternatively used after point-source removal.

In Figure \ref{vs100um}, we clearly detect the linear correlations between $S^{\rm diffuse}$ and $S_{\rm 100{\mu}m}^{\rm diffuse}$
at the large $S_{\rm 100{\mu}m}^{\rm diffuse}$, and more importantly, the expected breaks at $S_{\rm 100{\mu}m}^{\rm diffuse} \la 0.8$ 
MJy sr$^{-1}$ at the both IPP bands.
These inflection points are in very good agreement with the CIB brightness reported by \citet{lagache00}, 
$S_{\rm 100{\mu}m}^{\rm CIB} = 0.78 \pm 0.21$ MJy sr$^{-1}$, who made use of the same {\it COBE}/DIRBE measurements as used
for absolute calibration of the present 100 $\mu$m map.
However, we note that $S_{\rm 100{\mu}m}^{\rm CIB}$ here can include any brightness component which does not correlate 
with the interstellar dust emission.
Actually \citet{dole06} point out that the CIB brightness measured by \citet{lagache00} is likely to include the residual 
ZL of $S_{\rm 100{\mu}m}^{\rm resZL} \sim$ 0.3 MJy sr$^{-1}$.
The present results indicate that a similar amount of the residual ZL is present in the 100 $\mu$m map of \citet{sfd98}.

Our estimates of the COB brightness are derived as the $S^{\rm diffuse}$ values on the observed 
$S^{\rm diffuse}$--$S_{\rm 100{\mu}m}^{\rm diffuse}$ linear correlations made by the DGL, at the point where 100 $\mu$m brightness 
equals 
to the CIB (i.e., $S_{\rm 100{\mu}m}^{\rm diffuse} = S_{\rm 100{\mu}m}^{\rm CIB}$ in Equation (\ref{eq:dgl})).
First, we obtain the regression lines of the $S^{\rm diffuse}$--$S_{\rm 100{\mu}m}^{\rm diffuse}$ relations at 
$S_{\rm 100{\mu}m}^{\rm diffuse}$ = 1.0 -- 3.0 MJy sr$^{-1}$ using the least-$\chi^2$ method.
The derived mean DGL-to-100 $\mu$m brightness ratios are $a_d^0 = 3.2 \pm 0.1$ and $3.4 \pm 0.1$ bgu (MJy sr$^{-1}$)$^{-1}$
at $B_{\rm IPP}$ and $R_{\rm IPP}$, respectively.
Then the intersects of the regression lines and $S_{\rm 100{\mu}m}^{\rm diffuse} = S_{\rm 100{\mu}m}^{\rm CIB}$ are derived, 
whose $S^{\rm diffuse}$ values give our COB estimates.
Here, we adopt the CIB brightness reported by \citet{lagache00} ($S_{\rm 100{\mu}m}^{\rm CIB}$ = 0.78 $\pm$ 0.21 MJy sr$^{-1}$)
and include the associated uncertainty in the $S^{\rm COB}$ errors.
The results are $S^{\rm COB}$ = 1.8 $\pm$ 0.7 and 1.2 $\pm$ 0.7 bgu at $B_{\rm IPP}$ and $R_{\rm IPP}$, respectively.

%We obtain the regression lines of the observed $S^{\rm diffuse}$ -- $S_{\rm 100{\mu}m}^{\rm diffuse}$ relations at 
%$S_{\rm 100{\mu}m}^{\rm diffuse}$ = 1.0 -- 3.0 MJy sr$^{-1}$ using the least-$\chi^2$ method.
%The derived mean DGL-to-100 $\mu$m brightness ratios are $a_d^0 = 3.2 \pm 0.1$ and $3.4 \pm 0.1$ bgu/(MJr sr$^{-1}$) at $B_{\rm IPP}$ 
%and $R_{\rm IPP}$, respectively.
%Then the intersects of the regression lines and $S_{\rm 100{\mu}m}^{\rm diffuse} = S_{\rm 100{\mu}m}^{\rm CIB}$ are derived, 
%whose $S^{\rm diffuse}$ values give our estimates of the COB brightness.
%Here we adopt $S_{\rm 100{\mu}m}^{\rm CIB}$ = 0.78 $\pm$ 0.21 MJy sr$^{-1}$ \citep{lagache00} and include the associated uncertainty 
%in the $S^{\rm COB}$ errors.
%The results are $S^{\rm COB}$ = 1.8 $\pm$ 0.7 and 1.2 $\pm$ 0.7 bgu at $B_{\rm IPP}$ and $R_{\rm IPP}$, respectively.

Recently, \citet{penin11} present a CIB brightness of $S_{\rm 100{\mu}m}^{\rm CIB} = 0.20 \pm 0.09$ MJy sr$^{-1}$ from
the analysis of the {\it Spitzer Space Telescope} ({\it Spitzer}) Multiband Imaging Photometer for {\it Spitzer} (MIPS) data.
Based on the MIPS calibration at 160 $\mu$m, the 100 $\mu$m CIB is derived from the CIB anisotropy color between 
100 and 160 $\mu$m computed from the MIPS and the IRIS data.
The reason for the discrepancy between their result and the \citet{dole06} revision of the \citet{lagache00} result, $0.48 \pm 0.21$ 
MJy sr$^{-1}$, remains unclear.
However, the latter is the better choice for the present study, since it is based on the same DIRBE calibration as used for
the 100 $\mu$m map of \citet{sfd98}.
For reference, the estimated COB brightness would be reduced by $\sim$0.9 bgu when the result of \citet{penin11} is used with the 
plausible residual ZL level of $S_{\rm 100{\mu}m}^{\rm resZL} \sim$ 0.3 MJy sr$^{-1}$ in the 100 $\mu$m map (see above).

The above $S^{\rm COB}$ errors do not include associated systematic uncertainties. 
Median IPP brightness of the fields with $S_{\rm 100{\mu}m}^{\rm diffuse}$ = 0.78 $\pm$ 0.21 MJy sr$^{-1}$ {\it before} 
starlight subtraction is $S \sim$ 40 bgu at the both bands,
which leads to additional error of 0.4 bgu in the presence of 1\% systematic uncertainty in the IPP measurements
as estimated above.
The {\it Tycho-2} and GSC-II stars contribute about equally (i.e., $\sim$ 20 bgu each) to the above starlight while fainter stars 
make negligible contribution (see Section 5).
Taking 0.02 mag systematic uncertainties of the stellar magnitudes into account, 0.5 bgu error is further added.
After combining all these errors, the final estimates of the COB brightness are $S^{\rm COB}$ = 1.8 $\pm$ 0.9 
and 1.2 $\pm$ 0.9 bgu at $B_{\rm IPP}$ and $R_{\rm IPP}$, respectively, or 7.9 $\pm$ 4.0 and 7.7 $\pm$ 5.8 nW m$^{-2}$ sr$^{-1}$.
%The derived 1$\sigma$ confidence intervals are shown in Figure \ref{vs100um} by the shaded areas.
The results are summarized along with the derived $a_d^0$ values in Table \ref{tab:results}.

%On the other hand, the strict lower limits of the COB brightness have been obtained from the galaxy counts by the deep imaging observations.
%\citet{madau00} present the 1$\sigma$ lower limits of the integrated brightness of galaxies, 0.9 bgu at 0.45 $\mu$m and 0.67 $\mu$m,
%in the {\it Hubble} Deep Field (HDF).
%This result is confirmed by the similar analysis by \citet{totani01}.
%We combine the above lower limits with the estimates from the present analysis, which give our final estimates of the COB
%brightness: $S^{\rm COB} = 1.8^{+0.9}_{-0.9}$ and $1.2^{+0.9}_{-0.3}$ bgu at $B_{\rm IPP}$ and $R_{\rm IPP}$, respectively.
%They correspond to $7.9^{+4.0}_{-4.0}$ and $7.7^{+5.8}_{-1.9}$ nW m$^{-2}$ sr$^{-1}$.

\begin{table*}
\begin{center}
\caption{COB Brightness and Mean DGL-to-100 $\mu$m Brightness Ratios $a_d^0$ \label{tab:results}}
\begin{tabular}{cccccc}
\tableline\tableline
               & Wavelength  & \multicolumn{2}{c}{COB Brightness}         & \multicolumn{2}{c}{DGL-to-100 $\mu$m Ratio $a_d^0$}\\
    Band       & ($\mu$m)    &  (bgu)         & (nW m$^{-2}$ sr$^{-1}$)  & (bgu [MJy sr$^{-1}$]$^{-1}$) &  (dimensionless) \\
\tableline
 $B_{\rm IPP}$ & 0.44 & 1.8 $\pm$ 0.9 & 7.9 $\pm$ 4.0 & 3.2 $\pm$ 0.1 & (2.1 $\pm$ 0.1) $\times 10^{-3}$\\
 $R_{\rm IPP}$ & 0.64 & 1.2 $\pm$ 0.9 & 7.7 $\pm$ 5.8 & 3.4 $\pm$ 0.1 & (4.6 $\pm$ 0.1) $\times 10^{-3}$\\
\tableline
\end{tabular}
%% Any table notes must follow the \end{tabular} command.
%\tablenotetext{a}{Heliocentric distance of the spacecraft.}
%\tablecomments{The COB brightness can be converted to the \\
%  units of $\nu I_{\nu}$ (nW m$^{-2}$ sr$^{-1}$) by multiplying by 4.4 and \\
%  6.4 in the $B_{\rm IPP}$ and $R_{\rm IPP}$ band, respectively.}
\end{center}
\end{table*}

\section{Discussion \label{sec:discussion}}

\subsection{Correlation between the DGL and the 100 $\mu$m Emission}

%The obtained DGL-to-100 $\mu$m brightness ratios are $a_d$ = $3.2 \pm 0.1$ and $3.4 \pm 0.1$ bgu/(MJy sr$^{-1}$) in the
%$B_{\rm IPP}$ and $R_{\rm IPP}$ band, respectively.
By converting bgu unit to MJy sr$^{-1}$, we obtain dimensionless representations of the mean DGL-to-100 $\mu$m brightness 
ratios $a_d^0$ = $(2.1 \pm 0.1) \times 10^{-3}$ and $(4.6 \pm 0.1) \times 10^{-3}$ at
$B_{\rm IPP}$ and $R_{\rm IPP}$, respectively, which are useful for a comparison with other studies.
These values are also found in Table \ref{tab:results}.
The previous measurements of $a_d^0$ are compiled by \citet{bernstein02a} in their Table 7.

At the blue band around 0.45 $\mu$m, \citet{guhathakurta89} find the values of $a_d^0$ = $1.1 \times 10^{-3}$ and 
$2.6 \times 10^{-3}$ in their low 100 $\mu$m brightness regions "ir 2" and "ir 3".
They can be reproduced by a simple model of the interstellar dust scattering the ISRF, which predicts $a_d^0$ = $3.3 \times 10^{-3}$.
A similar value of $a_d^0$ = $1.1 \times 10^{-3}$ is observed by \citet{paley91}.
\citet{bernstein02a} provide another dust scattering model and show that $a_d^0$ = $1.6 \times 10^{-3}$ at the high 
Galactic latitudes.
From the observations of optically thin high Galactic latitude clouds, \citet{witt08} find the ratio of $a_d^0$ = $2.1 \times 10^{-3}$.
Overall, our result at $B_{\rm IPP}$ agrees with these previous measurements and model predictions.

At the red band around 0.65 $\mu$m, the measured $a_d^0$ values are $2.2 \times 10^{-3}$ and $4.4 \times 10^{-3}$ in
the ir 2 and ir 3 regions of \citet{guhathakurta89}.
\citet{paley91} report $a_d^0$ = $8.5 \times 10^{-3}$ at 0.55 $\mu$m and $11 \times 10^{-3}$ at 0.70 $\mu$m.
These measurements point the $a_d^0 (R)$/$a_d^0 (B)$ ratios of at least 2.0, while they are known to be significantly
larger than the predictions of dust scattering models.
In fact the models of \citet{guhathakurta89} and \citet{bernstein02a} consistently predict $a_d^0 (R)$/$a_d^0 (B)$ $\sim$ 1.4.
The excess component in the red band is called extended red emission (ERE) and is found in many Galactic objects including
reflection nebulae \citep{witt84,witt90}, infrared cirrus \citep{guhathakurta89}, and planetary nebulae \citep{furton90,furton92}.
Its origin is most likely the photoluminescence process of materials such as hydrogenated amorphous carbon and
polycyclic aromatic hydrocarbon excited by UV/optical photons \citep[see][for a review]{witt04}.
Our result, $a_d^0 (R)$/$a_d^0 (B)$ = 2.2, is consistent with the previous observations, suggesting that ERE is also present 
in the diffuse ISM with the lowest far-IR brightness.
It confirms the finding of \citet{gordon98} who reach the same conclusion from the analysis of the IPP data.

In summary, our results are in overall agreement with the previous observations toward the denser dust regions.
%However, the presented $a_d^0$ values should be preferred than the previous results in describing the diffuse ISM, since the latter 
%usually focus on the {\it enhanced} DGL components above background of the diffuse ISM emission.
A further study on this issue is beyond the scope of this work and will be presented in a future paper.

\subsection{Resolved Fraction of Cosmic Background}

\begin{figure*}
\epsscale{0.9}
\plotone{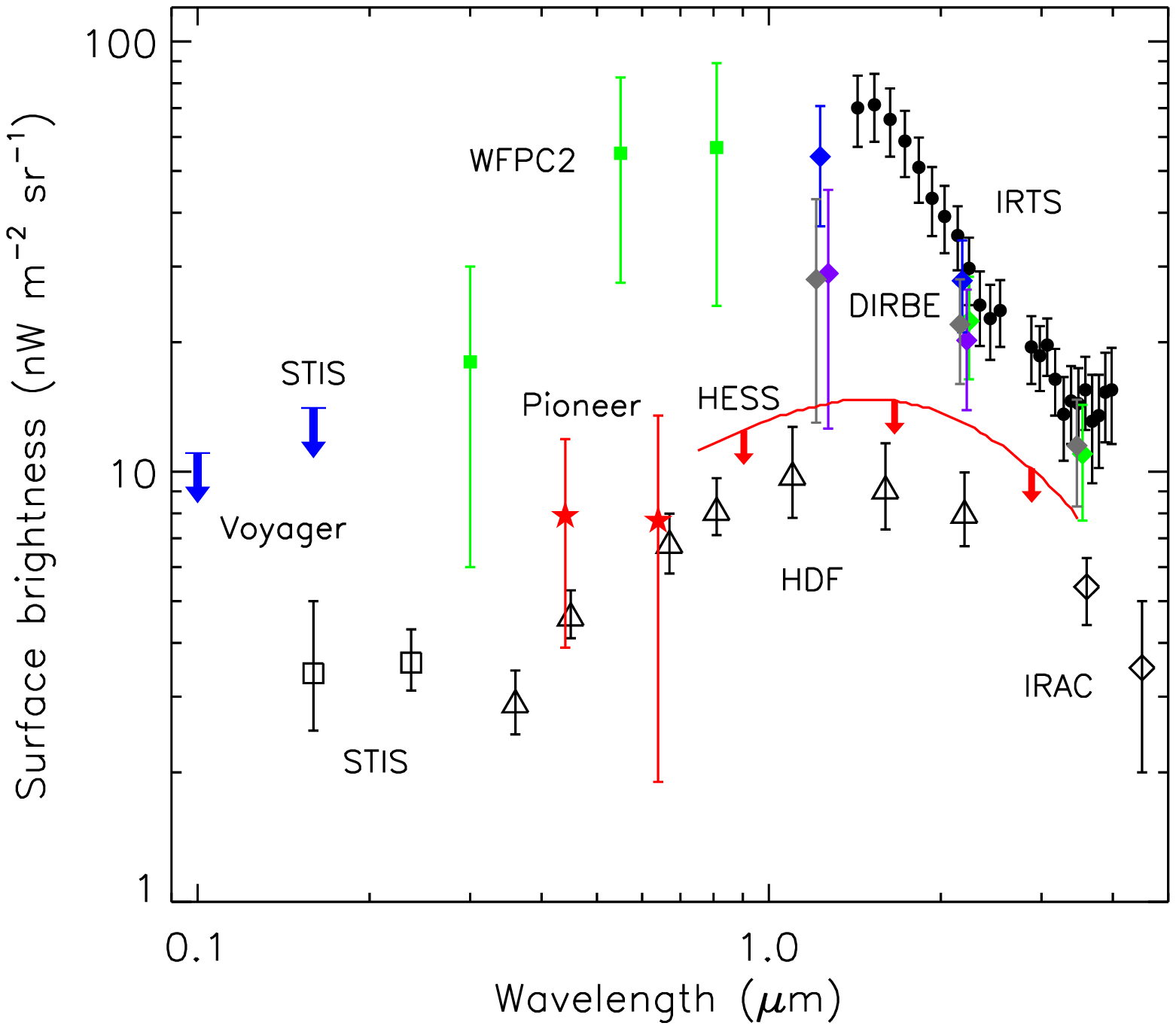}%{../../ebl_spec.ps}
\caption{Current measurements of the cosmic background (filled symbols) and the integrated brightness of galaxies 
  (open symbols) at UV, optical, and near-IR wavelengths.
  The cosmic background measurements include the UV upper limits (blue arrows) at 0.10 $\mu$m obtained from 
  the {\it Voyager}/UVS \citep{edelstein00} and at 0.16 $\mu$m from the {\it HST}/STIS \citep{brown00},
  the claimed detections at optical wavelengths using the {\it HST}/WFPC2 \citep[][green squares]{bernstein07}
  and at near-IR wavelengths using the {\it COBE}/DIRBE [\citet{gorjian00}, green diamonds; 
  \citet{wright01}, purple diamonds; \citet{cambresy01}, blue diamonds;  \citet{wright04}, grey diamonds;
  the wavelengths of these measurements are slightly shifted relative to each other for clarity]
  and the IRTS \citep[][black circles]{matsumoto05}.
  The red stars are the {\it Pioneer}/IPP results of this work, while the red solid line with arrows between 0.8 and 4 $\mu$m 
  represents the HESS upper limits \citep{aharonian06}.
  The integrated brightness of galaxies come from the {\it HST}/STIS measurements at UV \citep[][squares]{gardner00},
  the HDF compilation from UV to near-IR \citep[][triangles]{madau00}, and the {\it Spitzer}/IRAC measurements at 
  near-IR wavelengths \citep[][diamonds]{fazio04}.\label{ebl}}
\end{figure*}

We compile the current measurements of the cosmic background and the integrated brightness of galaxies at ultraviolet (UV),
optical, and near-IR wavelengths in Figure \ref{ebl}.
At UV wavelengths, the upper limits of the cosmic background are obtained from the analysis of the {\it Voyager} Ultraviolet 
Spectrometer (UVS) 
data \citep{edelstein00} and from the {\it HST}/Space Telescope Imaging Spectrograph (STIS) observations \citep{brown00}.
They are a few times the integrated brightness of galaxies measured by the {\it HST}/STIS \citep{gardner00}, still
leaving a large gap to be bridged.
The situation in near-IR wavelength range is much more controversial.
\citet{matsumoto05} claim detection of the strong near-IR CIB based on the Infrared Telescope in Space (IRTS) data.
Their CIB values are marginally consistent with the results from the {\it COBE}/DIRBE measurements reported by several authors
\citep{gorjian00, wright01, cambresy01, wright04} at 1.25, 2.2, and 3.3 $\mu$m.
The integrated brightness of galaxies at these wavelengths, as well as at the optical $UBVI$ bands, are derived from 
the {\it Hubble} Deep Field (HDF) data set by \citet{madau00}.
Those at the four bands of the {\it Spitzer} Infrared Array Camera (IRAC) are presented by \citet{fazio04}.
Using the HDF and the Subaru Deep Field (SDF) data, \citet{totani01} obtain the consistent results with \citet{madau00}.
They also find that 80\%--90\% of the total light from normal galaxies has already been resolved in the SDF $J$ and $K$ bands,
based on a galaxy evolution model taking into account various selection effects of observations.
Therefore, the large CIB excess found by \citet{matsumoto05} should be attributed to either some exotic radiation sources 
such as Population III stars or the residual ZL in the IRTS data.
Another constraint comes from the High Energy Stereoscopic System (HESS) $\gamma$-ray observations of the blazars at 
$z$ = 0.17--0.19 by \citet{aharonian06}, who find that the unexpectedly hard spectra of these objects cannot
be reproduced without assuming very low upper limits of the CIB.
Their upper limits are close to the integrated brightness of galaxies and are in sharp contrast with the previous
CIB estimates.

At optical wavelengths, \citet{bernstein02a} claim the "first detections" of the COB at 0.30, 0.55, and 0.80 $\mu$m
(see Section 1).
They used the Wide Field Planetary Camera 2 (WFPC2) on board the {\it HST} for absolute surface brightness measurements 
of the night sky, and a ground-based telescope for contemporaneous measurements of the ZL brightness.
However, a number of problems are raised by \citet{mattila03} for their analysis, resulting in a series of corrections of the 
presented results \citep{bernstein05, bernstein07}.
Their latest values are plotted in Figure \ref{ebl}, which are nearly one order of magnitude larger than the integrated 
brightness of galaxies and seem to be hard to explain in the current framework of galaxy evolution.
%Their original COB values are 4.0 $\pm$ 2.5, 2.7 $\pm$ 1.4, and 2.2 $\pm$ 1.0 bgu at the three wave bands.
%However, a number of problems are raised by \citet{mattila03} for their analysis.
%He shows that the corrected COB brightness should be 2 -- 7 times as high as the original values, which are clearly in
%conflict with the past estimates \citep[e.g.,][]{dube79,toller83}.
%After the initial correction in \citet{bernstein05}, \citet{bernstein07} present the significant revision of their original
%analysis, resulting in the COB brightness of 6 $\pm$ 4, 10 $\pm$ 5, and 7 $\pm$ 4 bgu (plotted in Figure \ref{ebl}).
%However, these latest values are nearly 1 order of magnitude larger than the integrated brightness of galaxies, which is
%hard to explain in the current framework of galaxy evolution.

Our new results of the COB brightness are mildly larger than and consistent within 1$\sigma$ uncertainty with the integrated
brightness of galaxies.
They are on the smooth extension of the upper limits found by \citet{aharonian06}.
With our best estimates, approximately 60\% and 90\% of the COB have already been resolved into discrete galaxies in the HDF
at 0.44 $\mu$m ($B_{\rm IPP}$) and 0.64 $\mu$m ($R_{\rm IPP}$), respectively.
On the other hand, \citet{totani01} demonstrate that 60\%--90\% and 80\%--100\% of the total light from galaxies have been
resolved at 0.45 and 0.61 $\mu$m.
The above facts indicate that bulk of the COB are comprised of normal galaxies, and there are little room for contributions of 
other populations at these wavelengths.
Note that AGNs are one of such populations, while their contributions to the cosmic background are estimated to be 
less than 10\%--20\% \citep{madau00}.

Finally, we comment on the possibility of the residual light from within the Galaxy in the detected COB.
In Section \ref{sec:mapgeneration}, we show that the effect of the residual ZL is little, since the systematic difference between 
the two data subsets taken at the systematically different heliocentric distances is no larger than those from other 
factors, which are actually included in our error budget.
It is consistent with the finding of \citet{hanner74} that there are no detectable ZL beyond $R$ = 3.3 AU.
While the starlight subtraction at the fainter magnitudes than the GSC-II detection limits rely on the star-count model down 
to 32.0 mag, absolute contributions of these faintest stars are less than $\sim$0.1 bgu at the both bands at the Galactic 
latitudes $|b| > 60^{\circ}$.
Therefore, their uncertainties cannot significantly affect our results.
Differential starlight brightness at the faint limit of 32.0 mag is less than 1 $\times$ 10$^{-5}$ bgu mag$^{-1}$, 
hence even fainter stars are not likely to make considerable contribution to the detected COB levels.
The DGL components which do not correlate with 100 $\mu$m emission are other sources of radiation that cannot 
be removed from the IPP brightness in the present analysis.
Line and continuum emissions from the interstellar gas might be candidates of such components, if their spatial distributions 
are significantly different from that of the interstellar dust producing the dominant component of the DGL.
However, no such emissions are currently known at the level of exceeding a few times 0.1 bgu (see Section \ref{sec:dgl}).
In any case, the residual light from within the Galaxy in the detected COB would indicate the {\it true} COB brightness even 
smaller than the present estimates, and hence further strengthen our conclusion that the bulk of the COB has already been 
resolved into galaxies by the current deepest observations.

\section{Summary \label{sec:summary}}

We present the results from an analysis of the {\it Pioneer 10/11} IPP measurements.
The used data were obtained when the spacecrafts were at the heliocentric distances $R > 3.3$ AU, where brightness of the
ZL is below the detection limit of the instruments.
We carefully examine the data quality and select about 70\% of the whole data as usable for the present analysis.
The selected data are integrated into sky maps of the high Galactic latitudes ($|b| > 35^{\circ}$) at the two bands of the
IPP, $B_{\rm IPP}$ ($\sim$0.44 $\mu$m) and $R_{\rm IPP}$ ($\sim$0.64 $\mu$m).
Systematic and random uncertainties of the measurements are estimated to be within 1\% and 2\%, respectively, at the both bands.

Galactic starlight in the IPP brightness is subtracted using the two all-sky star catalogs, namely, the {\it Tycho-2} catalog and 
the {\it HST} GSC-II.
Contribution of fainter stars than the detection limits of these catalogs are estimated from the stellar population synthesis
code TRILEGAL, while they are found to be negligible.
The DGL is further separated out from the IPP brightness by making use of its correlation with thermal 
100 $\mu$m emission taken from the {\it IRAS}--{\it COBE}/DIRBE map.
The final residual light represents the extragalactic component, i.e., the COB.

The derived COB brightness is 1.8 $\pm$ 0.9 and 1.2 $\pm$ 0.9 bgu at $B_{\rm IPP}$ and $R_{\rm IPP}$, respectively, 
or 7.9 $\pm$ 4.0 and 7.7 $\pm$ 5.8 nW m$^{-2}$ sr$^{-1}$.
From comparison with the integrated brightness of galaxies, we conclude that bulk of the COB is comprised of normal galaxies 
which have already been resolved in the current deepest observations.
On the other hand, the derived mean DGL-to-100 $\mu$m brightness ratios, $2.1 \times 10^{-3}$ and $4.6 \times 10^{-3}$ at the two
bands, are roughly consistent with the previous observations toward the denser dust regions.
The red color of these ratios confirm the presence of ERE in the diffuse ISM.

%The IPP brightness maps created in this work are available via the Astrophysical Journal website.\footnote{
%They are provided as the multi-extension fits files:
%"\texttt{IPP\_Bmap\_N/SGP.fits}" and "\texttt{IPP\_Rmap\_N/SGP.fits}" for the $B_{\rm IPP}$- and $R_{\rm IPP}$-band
%sky brightness in the north/south Galactic hemisphere, respectively.
%The description of each map is contained in the header unit.}

\acknowledgments

We are grateful to the referee for his/her useful comments.
This work was supported by Grants-in-Aid for Scientific Research (21840027, 22684005), Specially Promoted Research on 
Innovative Areas (22111503), and the Global COE Program of Nagoya University "Quest for Fundamental Principles in the Universe"
from JSPS and MEXT of Japan.

\end{document}